\documentclass[a4paper,11pt]{article}
\pdfoutput=1 % if your are submitting a pdflatex (i.e. if you have
\usepackage{jheppub} % for details on the use of the package, please
\usepackage{graphics} % Add graphics capabilities
\usepackage{amsmath}  % Better maths support
\usepackage{amsthm}
\usepackage{cancel}
\usepackage{ulem}

\usepackage[T1]{fontenc} % if needed

\renewcommand{\theequation}{\arabic{equation}}
\newcommand{\be}{\begin{equation}}
\newcommand{\ee}{\end{equation}}
\newcommand{\ea}{\end{array}}
\newcommand{\beqa}{\begin{eqnarray}}
\newcommand{\eeqa}{\end{eqnarray}}

\newcommand{\nn}{\nonumber}
\newcommand{\K}{k\!\!\!/}
\newcommand{\e}{\mathrm{e}}
\newcommand{\tr}{\mathrm{tr}}
\newcommand{\td}[1]{\tilde{#1}}
\newcommand{\al}{\alpha}
\newcommand{\bt}{\beta}
\newcommand{\dl}{\delta}
\newcommand{\g}{\gamma}
\newcommand{\s}{\sigma}
\newcommand{\ep}{\epsilon}

\newcommand{\pb}{\bar{p}}
\newcommand{\ph}{\hat{p}}
\newcommand{\gb}{\bar{g}}
\newcommand{\gh}{\hat{g}}
\newcommand{\zb}{\bar{z}}
\newcommand{\zh}{\hat{z}}
\newcommand{\wh}{\hat w}
\newcommand{\wb}{\bar w}
\newcommand{\p}{p\!\!\!/~}
\newcommand{\pbdash}{\bar{p} \!\!\!/~}
\newcommand{\q}{q\!\!\!/~}

\newtheorem*{theorem*}{Proposition}

\def\bear{\begin{eqnarray}}
\def\ear{\end{eqnarray}\noindent}
\def\bec{\blue\begin{equation}}
\def\eec{\end{equation}\black\noindent}
\def\bearc{\blue\begin{eqnarray}}
\def\earc{\end{eqnarray}\black\noindent}
\def\benn{\begin{enumerate}}
\def\enn{\end{enumerate}}

\def\mn{{\mu\nu}}
\def\Eins{{\mathchoice {\rm 1\mskip-4mu l} {\rm 1\mskip-4mu l}{\rm 1\mskip-4.5mu l} {\rm 1\mskip-5mu l}}}
\def\half{\frac{1}{ 2}}
\def\fourth{\frac{1}{4}}

\def\slash#1{#1\!\!\!\raise.15ex\hbox {/}}

\def\Gphat#1#2{{\hat{{\dot {{\cal G}}}}_{B#1#2}}}

\def\4piTD{{(4\pi T)}^{-{D\over 2}}}
\def\4piT4{{(4\pi T)}^{-2}}

\def\Tintm4{{\dps\int_{0}^{\infty}}{dT\over T}\,e^{-m^2T}
    {(4\pi T)}^{-2}}
\def\Tintm{{\dps\int_{0}^{\infty}}{dT\over T}\,e^{-m^2T}}

\newcommand{\slG}{{{\dot G}\!\!\!\! \raise.15ex\hbox {/}}}

\def\GBd12{{\dot G}_{B12}}

\newcommand{\no}{\noindent}
\def\non{\nonumber}
\def\dps{\displaystyle}
\def\sy{\scriptscriptstyle}
\def\sy{\scriptscriptstyle}
\def\partder#1#2{{\partial #1\over\partial #2}}
\def\td{\tilde d}

\def\no{\noindent}

\newcommand{\beq}{\begin{equation}}
\newcommand{\eeq}{\end{equation}}

\newcommand{\n}{{\mathfrak n}}

\def\i{\frak{iso}(1,3,\mathbb C)}
\def\>{\rangle}
\def\<{\langle}

\begin{document}

\title{Three-loop Euler-Heisenberg Lagrangian in 1+1 QED, part 1: single fermion-loop part}

%% %simple case: 2 authors, same institution
%% \author{A. Uthor}
%% \author{and A. Nother Author}
%% \affiliation{Institution,\\Address, Country}

% more complex case: 4 authors, 3 institutions, 2 footnotes
\author[a,b]{Idrish Huet,}
\author[c]{Michel Rausch de Traubenberg,}
\author[d,e]{Christian Schubert}
%\author[a,2]{and Fourth}

% The "\noindentte" macro will give a warning: "Ignoring empty anchor..."
% you can safely ignore it.

\affiliation[a]{Facultad de Ciencias en F\'isica y Matem\'aticas, Universidad Aut\'onoma de Chiapas\\
 Ciudad Universitaria, Tuxtla Guti\'errez 29050, Mexico}
\affiliation[b]{Theoretisch-Physikalisches Institut, Friedrich-Schiller-Universit\"at Jena,\\
Max-Wien-Platz 1, D-07743 Jena, Germany
}

\affiliation[c]{Universit\'e de Strasbourg, CNRS, IPHC UMR7178, F-67037 Strasbourg Cedex, France
}
\affiliation[d]{Instituto de F{{\'\i}}sica y Matem\'aticas, Universidad Michoacana de San Nicol\'as de Hidalgo\\
Apdo. Postal 2-82, C.P. 58040, Morelia, Michoacan, Mexico
}
\affiliation[e]
{Kavli Institute for Theoretical Physics,
University of California, 
Santa Barbara, CA  93106, USA.
}

% e-mail addresses: one for each author, in the same order as the authors
\emailAdd{idrish@ifm.umich.mx}
\emailAdd{Michel.Rausch@iphc.cnrs.fr}
\emailAdd{christianschubert137@gmail.com}

\abstract{
We study the three-loop Euler-Heisenberg Lagrangian in spinor quantum electrodynamics
in 1+1 dimensions. 
%This calculation is motivated mainly by the Affleck-Alvarez-Manton conjecture 
%concerning the exponentiation of the multi-loop corrections to Schwinger's formula for pair creation in a weak constant constant electric field.
In this first part we calculate the one-fermion-loop contribution, applying both standard
Feynman diagrams and the worldline formalism which leads to two different representations in terms of
fourfold Schwinger-parameter integrals. Unlike the diagram calculation, the worldline approach allows
one to combine the planar and the non-planar contributions to the Lagrangian.
Our main interest is in the asymptotic behaviour  of the weak-field expansion coefficients of this Lagrangian, for which a non-perturbative prediction has been
obtained in previous work using worldline instantons and Borel analysis. We develop algorithms for the
calculation of the weak-field expansions coefficients that, in principle, allow their calculation to arbitrary order.
Here for the non-planar contribution we make essential use of the polynomial invariants of the dihedral group $D_4$
in Schwinger parameter space to keep the expressions manageable. 
As expected on general grounds, the coefficients are of the form $r_1 + r_2 \zeta_3$ with rational
numbers $r_1,r_2$.
We compute the first two coefficients analytically, and four more by numerical integration.  
}

%\begin{document} 
\maketitle
\flushbottom

\section{Introduction}
\label{sec:intro}
\renewcommand{\theequation}{1.\arabic{equation}}
\setcounter{equation}{0}

In one of the first serious computations of early QED, Heisenberg and Euler in 1936
obtained their famous effective Lagrangian induced for a constant electromagnetic field
by an electron loop \cite{eulhei}.
For this effective Lagrangian they obtained the following one-parameter integral representation: 

\bear
{\cal L}^{(1)}(F) &=& - \frac{1}{8\pi^2}
\int_0^{\infty}{dT\over T^3}
\,\e^{-m^2T} 
\biggl[
{(eaT)(ebT)\over {\rm tanh} (eaT){\rm tan} (ebT)} 
%\nonumber\\&&\hspace{110pt}
- {1\over 3}(a^2-b^2)T^2 -1
\biggr]
\, .
\nonumber\\
\label{eulhei}
\ear
%(the superscript refers to the loop order).
Here $m$ and $T$ are the mass and proper-time of the electron and $a,b$  the Maxwell invariants, related to the
electric and magnetic fields by  $a^2-b^2 = B^2-E^2,\quad  ab = {\bf E}\cdot {\bf B}$. The second and third
term in brackets implement the renormalization of vacuum energy and charge. The superscript `(1)' stands for
one-loop.

\begin{figure}[htbp]
\begin{center}
\includegraphics[scale=1.1]{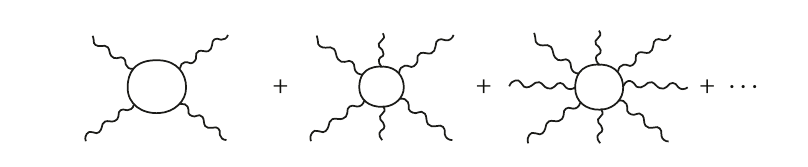}
%\caption{{\bf }}
\end{center}
\caption{Sum of diagrams equivalent to the one-loop EHL.}
\label{fig-eh1loopexpand}
\end{figure}

This Euler-Heisenberg Lagrangian (`EHL' in the following) is to be added to the classical
Maxwell Lagrangian, and contains a wealth of information on nonlinear quantum effects such as the
field-dependence of the speed of light, vacuum birefringence, and dichroism
(see \cite{ditgie-book,dunnerev} for reviews). Moreover, it contains this information in a form which is ready for
use with standard methods of nonlinear optics. 
From a particle theory point of view, the EHL holds the information on the 
one-loop $N$ - photon amplitudes for arbitrary $N$ in the low-energy limit
(see, e.g., \cite{itzzub-book,dunnerev}).   
In terms of Feynman diagrams, it is thus 
equivalent to the sum of graphs shown in Fig. \ref{fig-eh1loopexpand},
where all photon energies are small
compared to the electron mass, $\omega_i\ll m$.

\bigskip

The construction of  these low-energy amplitudes from the weak-field expansion of the EHL, 

\bear
{\cal L} (F) = \sum_{k,l} c_{kl}\, a^{2k}b^{2l} \, .
\label{wfe}
\ear
\noindent
can, using standard spinor helicity techniques, be carried out explicitly for any number
of photons and helicity assignments \cite{56,118}.
It turns out that, in this limit, once a helicity assignment has been fixed for all
photons, the full dependence of the $N$ - photon amplitude 
on the external momenta and polarization vectors can be absorbed into
a single invariant $\chi_N$. 

Except for the purely magnetic case, the EHL has also an imaginary part related
to vacuum pair creation by the electric field component  \cite{schwinger51}. 
In the purely electric case, which we will focus on here for simplicity, this imaginary part allows the 
following decomposition due to Schwinger \cite{schwinger51}:

\begin{eqnarray}
{\rm Im} {\cal L}^{(1)}(E) &=&  \frac{m^4}{8\pi^3}
\beta^2\, \sum_{k=1}^\infty \frac{1}{k^2}
\,\exp\left[-\frac{\pi k}{\beta}\right] 
\nonumber\\
\label{schwinger}
\end{eqnarray}
($\beta = eE/m^2$). 
Physically, the $k$th term in this decomposition relates to the coherent production of $k$ electron-positron pairs
in one Compton volume of spacetime (see, e.g., \cite{ritus-ginzburg}). 
The nonperturbative dependence of these ``Schwinger exponentials'' on the field supports the interpretation
of field-induced pair creation as a vacuum tunneling effect, as envisioned by Sauter
as early as 1931 \cite{sauter}.

The $k$th term in the decomposition (\ref{schwinger}) can be obtained from the $k$th pole in the proper-time integrand
of (\ref{eulhei}) by a simple application of the residue theorem. 
Still, for higher-loop considerations it turns out to be advantageous to observe that the
real and imaginary parts of the EHL can be related alternatively through the asymptotic properties 
of the coefficients of the weak-field expansion (\ref{wfe}) \cite{37}.
In the purely electric case, this expansion is 

\bear
{\cal L}^{(1)}(E)&=& - {1\over 8\pi^2}
\int_0^{\infty}{dT\over T^3}
\,\e^{-m^2T} 
\Biggl[
{eET\over \tan(eET)} + {1\over 3}(eET)^2 -1
\Biggr]
\nonumber\\
&=&
\frac{2m^4}{\pi^2}
\sum_{n=2}^{\infty}
c_n^{(1)}g^n
\label{EHLel}
\ear
where $g\equiv \Bigl({eE\over m^2}\Bigr)^2$ and

\bear
c_n^{(1)} = (-1)^{n+1}{2^{2n-4}B_{2n}\over (2n)(2n-1)(2n-2)} 
\label{cnBernoulli}
\ear
with $B_N$ the Bernoulli numbers. Replacing the weak-field expansion coefficients
$c_n^{(1)}$ by their leading asymptotic growth, which is

\bear
c_n^{(1)} \sim  {1\over 8}{\Gamma(2n-2)\over\pi^{2n}}
%\nonumber\\
%c_n^{(1)} \sim (-1)^n{1\over 8}{\Gamma(2n-2)\over\pi^{2n}}
%{1\over 2^{2n}}
%&\rightarrow&
%{\rm Im}\,{\cal L}^{(1)}(E) \sim {m^4\over 8\pi^3}
%\biggl({eE\over m^2}\biggr)^2{1\over 2^2}{\rm exp}\biggl(-2{\pi m^2\over eE}\biggr) \, ,
%\nonumber\\
%\vdots && \vdots
\nonumber\\
\label{match}
\ear
one can Borel sum the series. One obtains a Borel integral that is singular, leading to an imaginary part
that is precisely the leading term in the Schwinger decomposition (\ref{schwinger}):

\bear
{\rm Im}\,{\cal L}^{(1)}(E) 
\,\,\,\, {\stackrel{\beta\to 0}{\sim}} \,\,\,\,
{m^4\over 8\pi^3}
\biggl({eE\over m^2}\biggr)^2{\rm exp}\biggl(-{\pi m^2\over eE}\biggr) \, .
\label{summed}
\ear
This procedure can be repeated with the subleading, sub-subleading etc. asymptotic growth of the coefficients
$c_n^{(1)}$ to reproduce the full Schwinger expansion in (\ref{schwinger}) \cite{37}. 
However, in this paper we will be concerned only with the leading term in this expansion,
which dominates in the weak-field limit. 

\begin{figure}[h]
%\begin{picture}(0,0)(6000,10000)
{\centering
\hspace{70pt}\includegraphics[scale=1.1]{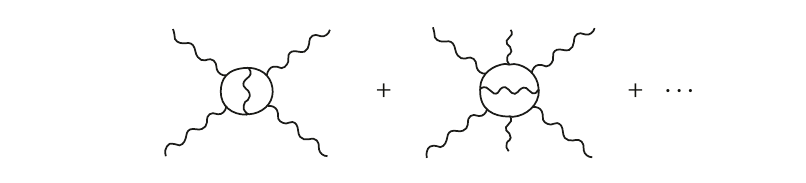}
}
\caption{Feynman diagrams contributing to the 2-loop EHL.}
\label{fig-eh2loopexpand}
\end{figure}

Our interest here is in multi-loop corrections to the EHL. 
The two-loop correction to the EHL, involving one internal photon exchange, 
corresponds to the diagrams shown in Fig. \ref{fig-eh2loopexpand} 
(here it is understood that internal photon corrections are  put in all possible ways). 

\bigskip

This two-loop EHL was first studied by Ritus in 1975 \cite{ritusspin}. That calculation, as well as later recalculations 
\cite{ditreu-book,18,24}, led to a representation of ${\cal L}^{(2)}$ in terms of rather  intractable two-parameter integrals, 
and so far only the first few coefficients of its weak-field expansion have been computed \cite{ritusspin,18,korsch,37,66}. 

Schwinger's formula for the imaginary part (\ref{schwinger}) also generalizes to the two-loop level, in the following way \cite{ritusspin,lebrit,ritus-ginzburg}:

\begin{eqnarray}
{\rm Im} {\cal L}^{(2)} (E) &=&  \frac{m^4}{8\pi^3}
\beta^2\,
\sum_{k=1}^\infty
\alpha\pi K_k^{\rm spin}(\beta)
\,\exp\left[-\frac{\pi k}{\beta}\right] 
\nonumber\\
%{\rm Im} {\cal L}_{\rm scal}^{(2)} (E) &=&  \frac{m^4}{16\pi^3}
%\beta^2\,
%\sum_{k=1}^\infty
%(-1)^{k+1}
%\alpha\pi K_k^{\rm scal}(\beta)
%\,\exp\left[-\frac{\pi k}{\beta}\right] \, ,
%\nonumber\\
\label{ImL2}
\end{eqnarray}
($\alpha=\frac{e^2}{4\pi}$).
%\begin{eqnarray}
%K_k^{\rm scal,spin}(\beta) &=& -{c_k\over \sqrt{\beta}} + 1 + {\rm O}(\sqrt{\beta}) \, ,
%\nonumber\\
%c_1 = 0,\quad && \quad
%c_k = {1\over 2\sqrt{k}}
%\sum_{l=1}^{k-1} {1\over \sqrt{l(k-l)}},
%\quad k \geq 2 \, ,
%\label{expK}
%\end{eqnarray}
%(these coefficients, also called $c_n$, should not be confused with the weak-field expansion coefficients
%introduced above).
Thus at two-loop one finds the same decomposition into Schwinger exponentials, but
the prefactor of the $k$th Schwinger-exponential now is not a constant but a function 
 $K_k (\beta)$ of the field strength. 
All that is presently known about these functions $K_k (\beta)$ explicitly are their
leading orders in the weak-field epxansion \cite{ritusspin,lebrit}. At this leading order,
things become extremely simple: adding up the one-loop and two-loop EHL's one finds 
  
\begin{eqnarray}
{\rm Im} {\cal L}^{(1)} (E) +
{\rm Im}{\cal L}^{(2)} (E) 
\,\,\,\, {\stackrel{\beta\to 0}{\sim}} \,\,\,\,
 \frac{m^4\beta^2}{8\pi^3}
\bigl(1+\alpha\pi\bigr)
\,{\rm e}^{-{\pi\over\beta}} \, .
\label{Im1plus2}
\end{eqnarray}
And in \cite{37} it was checked, albeit only numerically and based on a calculation
of fifteen coefficients, that the method of constructing the imaginary part of the    
EHL by Borel summation of the weak-field expansion coefficients still works fine
at the two-loop level, at least in this weak-field limit. However, it works now in a 
more complicated manner, since at the two-loop level mass renormalization kicks in. 
It turns out that, before mass renormalization, the coefficients of ${\cal L}^{(2)}$ 
have an asymptotic growth that is faster than what we have seen above for the one-loop
coefficients; only after adding the counterterm from mass renormalization, which is of the form
$\delta m \partder{}{m}{\cal L}^{(1)}$, and only if $\delta m$ is the physically renormalized mass (renormalized at the one-loop level) 
the addition of the counterterm leads to a cancellation that reduces the leading asymptotic growth of the two-loop coefficients to be precisely the same as at one loop. 

Thus the study of the two-loop EHL allows one to understand already two important aspects of the QED S-matrix:
first, although the physically renormalized electron mass is usually determined through a 
calculation of the electron propagator, it can as well be obtained from the effective
Lagrangian (although one has to go one loop higher in the perturbative expansion). 
This has been called ``immanent renormalization principle'' by Ritus (see, e.g., \cite{ritus-ginzburg}). 
Second, unless the physically renormalized electron mass is used, the two-loop $N$ - photon amplitudes dominate over the one-loop ones for sufficiently
large $N$, i.e. perturbation theory breaks down already at the two-loop level.

In \cite{lebrit} it was further noted that, if one would assume that in this 
weak-field approximation the higher-loop order corrections just lead to an exponentiation,

\begin{eqnarray}
\sum_{l=1}^{\infty}
{\rm Im} {\cal L}^{(l)} (E) \,\,\,\, {\stackrel{\beta\to 0}{\sim}} \,\,\,\,
{\rm Im} {\cal L}^{(1)} (E) 
\,{\rm e}^{\alpha\pi}
\label{Imall}
\end{eqnarray} 
then the ${\rm e}^{\alpha\pi}$ factor can be absorbed into the Schwinger factor ${\rm e}^{-{\pi\over\beta}}$
by the following mass-shift,

\bear
m(E) \approx m - \frac{\alpha}{2}\frac{eE}{m} \, .
\ear
This extrapolation, to be called ``exponentiation conjecture'' in the following, would appear far-fetched if based only on a two-loop calculation, but there is strong independent
support for its correctness:
first, the same mass shift had been found by Ritus already before 
for the crossed process of electron propagation in a constant electric field \cite{ritusmass}.
Second, in the vacuum tunneling picture of Sauter-Schwinger pair creation
it can be interpreted as taking into account the leading-order 
Coulomb energy correction due to the fact that the electron - positron
pair materializes at a finite distance \cite{lebrit}.
Third, the analogue exponentiation has also been obtained for Scalar QED by
Affleck et al. \cite{afalma}, and using a totally different approach based 
on a semi-classical approximation to Feynman's {\it worldline path integral presentation} of the multiloop QED effective action \cite{feynman50,feynman51}.

In 2002 G.V. Dunne and one of the authors \cite{50,51,52} noted that the electric or magnetic backgrounds are by no means the simplest ones
in the Euler-Heisenberg context. 
Computationally, the most favorable case is the one of a (euclidean)  self-dual (`SD') field.
Such a field is defined by $F=\tilde F$, which has the consequence that $F^2 = -f^2\Eins$.
For real $f$, the SD effective Lagrangian has properties similar to the magnetic EHL, for imaginary $f$ similar to the electric one. 
Surprisingly, for such a background the EHL turned out to be computable
in closed form even at the two-loop level:

\bear
%{\cal L}_{\rm scal}^{(2)(SD)}(\kappa)
%&=&
%\alpha \,{m^4\over (4\pi)^3}\frac{1}{\kappa^2}\left[
%{3\over 2}\xi^2 (\kappa)
%-\xi'(\kappa)\right]
%\label{L24Dscal}\\
{\cal L}^{(2)(SD)}(\kappa)
&=&
-2\alpha \,{m^4\over (4\pi)^3}\frac{1}{\kappa^2}\left[
3\xi^2 (\kappa)
-\xi'(\kappa)\right]
\label{L24Dspin}
\ear
where 

\begin{eqnarray}
\kappa\equiv \frac{m^2}{2ef}
\label{kappa}
\end{eqnarray}
and 

\bear
\xi(x)\equiv -x\Bigl(\psi(x)-\ln(x)+{1\over 2x}\Bigr)
\label{defxi}
\ear
with $\psi$ the digamma function $\psi(x)=\Gamma^\prime(x)/\Gamma(x)$.

This self-dual EHL cannot be realized with real fields in Minkowski space,
but still holds physical information on the photon amplitudes, namely on their
``all $+$'' helicity components  \cite{dufish1,dufish2}. 

Moreover, in the self-dual case the study of the weak-field expansions and the
construction of the imaginary parts involve only well-known properties
of the digamma function, and thus could be done much more
completely and explicitly than for the electric or magnetic case \cite{52}.
This made it possible to verify that the above-mentioned construction of the imaginary part by
Borel summation works as well at the two-loop level, that is, certain a priori possible complications
such as the appearance of additional poles or cuts in the complex $g$ plane apparently do not occur in QED.

Having thus gained confidence that the correspondence between the leading asymptotic behaviour of the weak-field
expansion coefficients (\ref{match}) and the leading weak-field behaviour of the imaginary part (\ref{summed}) of the EHL
persists to higher loop orders, it was suggestive to apply this correspondence in reverse to the
exponentiation conjecture to gain information on the multiloop photon amplitudes.  
This was carried out by G.V. Dunne and one of the authors in \cite{60}. 
There we essentially transferred the factor $\e^{\alpha \pi}$ first from the leading weak-field behaviour of the imaginary part to the leading large $N$ - behaviour of the expansion coefficients, and from there to the $N$ - photon low-energy amplitudes in the limit of $N\to\infty$. For the second step it was essential that, as mentioned above, in the low-energy limit the whole kinematical dependence of the photon amplitudes can be absorbed into a single
invariant, effectively reducing the amplitude to a single number. 

However, this leads to an analytic dependence on $\alpha$, which seems at variance with 
well-known arguments \cite{dyson,thooftborel} that exclude 
a non-vanishing radius of convergence for generic amplitudes in QED. 
For a further discussion of this apparent paradox see the recent \cite{111}. 

Considering its relevance for the asymptotic structure of the QED perturbation theory,
as well as its relation to the vacuum tunneling picture of Schwinger pair creation and
the Ritus mass shift, we consider it of the utmost importance to clarify whether the
exponentiation really works beyond the two-loop level. However, 
a calculation of the three-loop EHL seems presently technically out of reach
even for the simplest case of a self-dual field. 

Now, so far we have focused on spinor QED in $D=4$ dimensions. 
However, it is important to stress at this point, that spin has not played any role in the above issue. All the statements that we made above about the 
weak-field expansions at one and two loops hold, qualitatively, as well for Scalar QED \cite{ritusscal,18}, starting with
Weisskopf's ``Scalar EHL'' \cite{weisskopf}, 

\bear
{\cal L}_{\rm scal}^{(1)}(F)&=&  {1\over 16\pi^2}
\int_0^{\infty}{dT\over T^3}
\,\e^{-m^2T}
\biggl[
{(eaT)(ebT)\over \sinh(eaT)\sin(ebT)} 
%\nonumber\\&&\hspace{60pt}
+{e^2\over 6}(a^2-b^2)T^2 -1
\biggr]
\, .
\nonumber\\
\label{ehscal}
\ear
%For simplicity, we will call this Lagrangian `%in the following. The Schwinger-decomposition of the imaginary part differs from the
%spinor case, apart from the normalization, only by signs:
%
%\begin{eqnarray}
%{\rm Im}{\cal L}_{\rm scal}^{(1)}(E) 
%&=&
%-\frac{m^4}{16\pi^3}
%\beta^2\, \sum_{k=1}^\infty \frac{(-1)^{k}}{k^2}
%\,\exp\left[-\frac{\pi k}{\beta}\right] 
%\nonumber\\
%\label{schwingerscal}
%\end{eqnarray}
%Those sign changes incorporate the effect of Bose vs. Fermi statistics that kicks in for  multiple pair production. 
Similarly, at least at the one-loop level the structure of the EHL's and associated
Schwinger exponentials at one-loop is essentially independent of the
space-time dimension \cite{blviwi,gitgav}. In particular, in $D = 1+1$ dimensions the EHLs become

\bear
{\cal L}^{(1)(2D)}(f)&=& -{ef\over 4\pi}
\int_0^{\infty}{dZ\over Z}\e^{-2\kappa Z}
\Bigl(\coth(Z) - {1\over Z}\Bigr) 
\, ,
\label{L1spin}
\\
{\cal L}_{\rm scal}^{(1)(2D)}(f) &=& {ef\over 4\pi}
\int_0^{\infty}{dZ\over Z}\e^{-2\kappa Z}
\Bigl(\frac{1}{\sinh(Z)} - {1\over Z}\Bigr) 
\, ,
\label{L1scal}
\ear
where $Z= efT$, $\kappa = m^2/(2ef)$, and the constant field strength parameter $f$ is defined by

\bear
F &=& \left(\begin{array}{cc}0 & f \\-f & 0 \end{array}\right)
\label{defF}
\ear
 (see app. \ref{appconv} for our  conventions). We note that the integrals can also be done in closed form, e.g., for the
 Spinor QED case:
 
\bear
{\cal L}^{(1)(2D)}(\kappa ) &=& -{m^2\over 4\pi} {1\over\kappa}
\Bigl[{\rm ln}\Gamma(\kappa) - \kappa(\ln \kappa -1) +
\half \ln \bigl({\kappa\over 2\pi}\bigr)\Bigr]
\, .
\label{L1expl}
\ear

In 2006 Dunne and Krasnansky studied the Scalar QED EHL
in various dimensions at the two-loop level \cite{dunkra,krasnansky} and 
found, in particular, the following explicit formula for this EHL in the
$D=1+1$ case:

\bear
{\cal L}_{\rm scal}^{(2)(2D)}(\kappa)
&=&
-\frac{m^2}{2\pi} \frac{\tilde\alpha}{32}
\left[
\xi^2_{2D} 
-4\kappa \xi_{2D}'\right] 
\, .
\label{L22Dscal}
\ear
Here $\tilde\alpha \equiv 2\frac{e^2}{\pi m^2}$ is our definition of the fine structure constant
in two dimensions, and

\bear
\xi_{2D}(\kappa)&\equiv& -\Bigl(\psi(\kappa+\half)-\ln (\kappa)\Bigr)
=  \psi(\kappa)-2\psi(2\kappa) + \ln (4\kappa)
\, .
\non\\
\label{defxi2D}
\ear
This result is formally very similar to the one above for the $D=4$ EHL in a self-dual background, eq. (\ref{L24Dspin}),
which suggests that QED in two dimensions (scalar or spinor) may be sufficiently similar to serve as a simpler testing ground for the 
exponentiation conjecture. 

But f1irst it had to be established whether the exponentiation conjecture itself admits a generalization to the two-dimensional case.
This was settled in \cite{81} where, using the same method of worldline instantons as Affleck et al, it was found that the exponentiation
 formula (\ref{Imall}) generalizes to 2D Scalar QED in the form

\bear
{\rm Im}{\cal L}^{({\rm all-loop})(2D)}_{\rm scal}(E)
\,\,
&{\stackrel{E\to 0}{\sim}}&
\,\,
 \frac{eE}{4\pi}
 \e^{-\frac{m^2\pi}{eE} - \frac{\pi}{2}\frac{m^2}{E^2}}
=
 \frac{eE}{4\pi}
 \e^{-\frac{m^2\pi}{eE} + \tilde\alpha \pi^2  \kappa^2}
 \, .
 \nonumber\\
\label{ImLallloop2D}
\ear
By Borel analysis, this leads to the following
formula for the limits of ratios of $l$ - loop to one - loop coefficients:

\bear
{{\rm lim}_{n\to\infty}} {c^{(l)}_{2D}(n)\over c^{(1)}_{2D}(n+l-1)} 
&=& {(\tilde\alpha\pi^2)^{l-1}\over (l-1)!} 
\label{AAM2Dcoeff}
\ear
where the expansion coefficients in 2D are defined as

\bear
{\cal L}^{(l)(2D)}(\kappa) &=& \frac{m^2}{2\pi} \sum_{n=1}^{\infty}(-1)^{l-1}c_{2D}^{(l)}(n) (i\kappa)^{-2n}
\, .
\label{defcn}
\ear
Thus, differently from the 4D case, in 2D the asymptotic growth of the weak-field expansion coefficients
increases with increasing loop level, which presumably is related to the fact that the Coulomb interaction
in two dimensions is confining. 
In \cite{81} it was further shown that the coefficients of the two-loop Scalar QED EHL found by Dunne and
Krasnansky, eq. (\ref{L22Dscal}), indeed fulfill the asymptotic prediction (\ref{AAM2Dcoeff}) with $l=2$.
And there also the corresponding Spinor QED calculation was performed, leading to 

\bear
{\cal L}^{(2)}(\kappa) &=& {m^2\over 4\pi}\frac{\tilde\alpha}{4}
\Bigl[-\xi_{2D}'(\kappa)
+\ln(\lambda_0 m^2) + \gamma + 2 \Bigr]
\, .
\label{L2spinwlfin}
\ear
%{\cal L}^{(2)}(f)
 %-{e^2\over 8\pi^2}\xi'(\kappa)
%= {e^2\over 8\pi^2}(\tilde \psi(\kappa) + \kappa \tilde\psi'(\kappa))
%\nonumber\\
%\label{L2spinwlfin}
%\ear
Here $\xi_{2D}$ is the same function that appeared in the Scalar QED case, eq.(\ref{defxi2D}),
and $\lambda_0$ an infrared cutoff that affects only the irrelevant vacuum term. 
Note that in 2D the Spinor QED EHL has a 
simpler structure than the Scalar QED one, since it involves the function $\xi_{2D}$ 
only linearly. Nevertheless, it is easy to check that the resulting weak-field
expansion coefficients obey the same asymptotic relation (\ref{AAM2Dcoeff}).
Thus in 2D QED we have an asymptotic prediction for the spinless case, and explicit
two-loop computation confirms this prediction and suggests that it does not depend on spin.
Moreover, the calculations were significantly simpler than in four dimensions, 
in particular not requiring mass renormalization; although in 2D QED (finite) mass renormalization exists, it turns out that, differently from the 4D case, the corresponding terms in the EHL do not contribute to the leading order growth of the weak-field expansion
coefficients, and thus also not to the weak-field limit of the imaginary part.  
All this provides strong motivation for pushing the calculation of the EHL in 2D to higher orders.

\begin{figure}[h]
\begin{center}
\includegraphics[angle=0, width=0.2\textwidth]{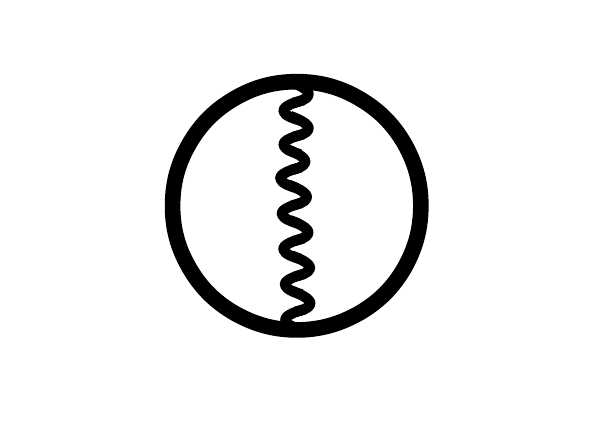}
\end{center}
\vspace{-20pt}
\caption{1PI diagram contributing to the two-loop EHL.}
\label{fig-EHL2loopIRR}
\end{figure}

\begin{figure}[h]
\begin{center}
\includegraphics[angle=0, width=0.5\textwidth]{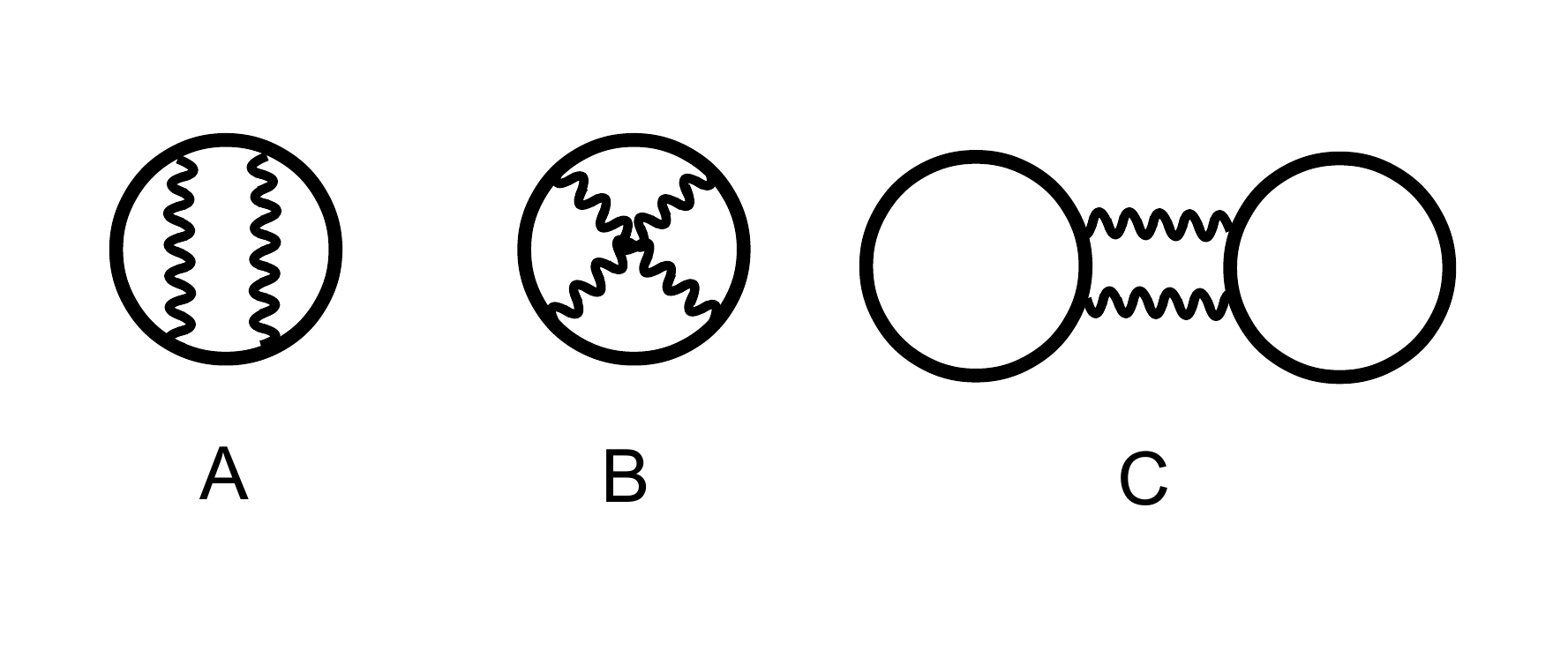}
\end{center}
\vspace{-20pt}
\caption{1PI diagrams contributing to the three-loop EHL.}
\label{fig-ABC}
\end{figure}

In this series of papers, we will present a complete calculation of the three-loop EHL in 2D spinor QED.
Until recently,  it was believed that this effective Lagrangian is (in any dimension) given at the two-loop
level by the diagram shown in Fig. \ref{fig-EHL2loopIRR}, and at the three-loop level by the three
diagrams shown in Fig. \ref{fig-ABC}. The electron propagators are the full ones in the external field.

\begin{figure}[h]
\begin{center}
\includegraphics[angle=0, width=0.2\textwidth]{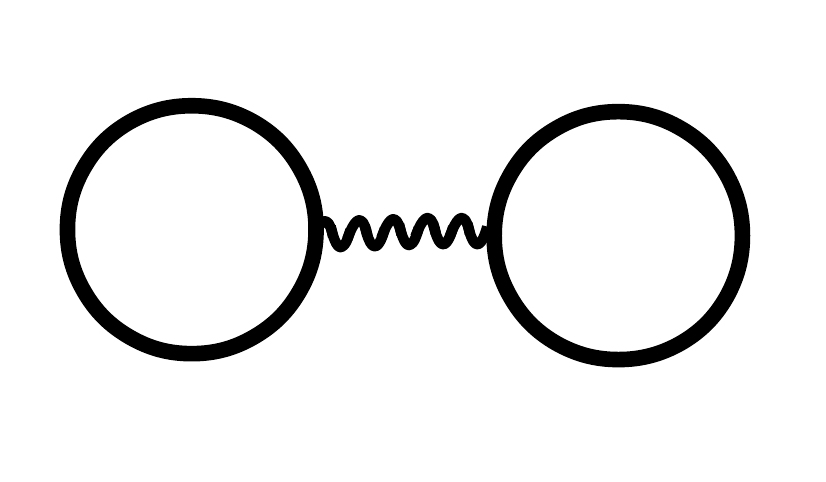}
\end{center}
\vspace{-20pt}
\caption{Reducible diagram contributing to the two-loop EHL.}
\label{fig-EHL2loopRED}
\end{figure}

\begin{figure}[h]
\begin{center}
\includegraphics[angle=0, width=0.4\textwidth]{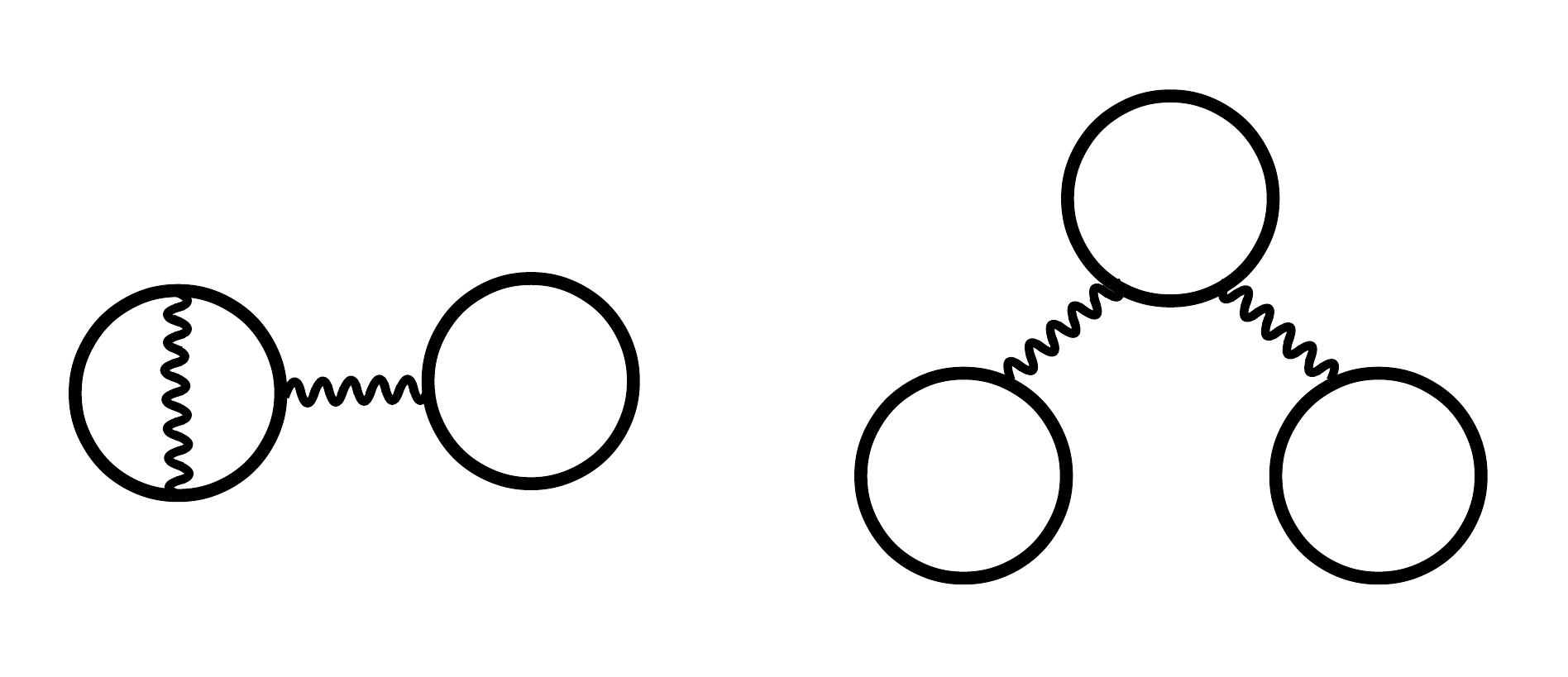}
\end{center}
\caption{Reducible diagrams contributing to the three-loop EHL.}
\label{fig-EHL3loopRED}
\end{figure}

However, in 2016 Gies and Karbstein \cite{giekar} (see also \cite{karbstein,112,113}) showed, that also the one-particle reducible diagrams shown in Figs.  
\ref{fig-EHL2loopRED} and \ref{fig-EHL3loopRED} have to be taken into consideration. 

Previously this type of diagram was believed to vanish, because it contains 
the one-loop tadpole diagram, which can be shown to formally vanish using momentum conservation and gauge invariance (see, e.g., \cite{ditreu-book,frgish-book}).
However, in \cite{giekar} it was shown that this is not the case any more when it appears as a subdiagram, due to the IR divergence of the photon propagator 
connecting it to the rest of the diagram in the zero - momentum limit. Thus we will have to include these diagrams here, too. 

In this first part of our three-loop analysis we will, however, restrict ourselves to the one-fermion loop (or ``quenched'') three-loop diagrams, that is, the diagrams
A and B of Fig. \ref{fig-ABC}. These are also the ones most relevant to our main object, the verification of the exponentiation conjecture at the three-loop level.

Partial results of this calculation have been published in various conference proceedings \cite{77,85,108,111,118}. 
A first attempt at this calculation \cite{77} used the standard approach to this type of calculation, namely Feynman diagrams with the exact electron propagator in the
field. The photon propagator was taken in Feynman gauge. Due to the super-renormalizability of 2D QED, at the three-loop level the effective Lagrangian is already
UV finite. However, we encountered spurious IR divergences that greatly complicated an already cumbersome calculation. 
In a second run \cite{85} we used the worldline formalism \cite{berkos-prl,berkos-npb,strassler1,strassler2}
along the lines of \cite{shaisultanov,18,41,41,91}, and also encountered IR divergences.
Those could be removed by suitable integrations by parts, but it then was found that it is also possible to avoid the appearance of IR divergences altogether by the
choice of a particular covariant gauge. This is the gauge $\xi = - 1$, to be called ``traceless gauge'' in the following, since it makes the photon propagator
traceless in $D=2$ (it had already been used in \cite{81}, but only in the worldline instanton calculation).
Returning then to the Feynman diagram calculation, we found that here, too, this gauge makes the IR finiteness manifest, and moreover leads
to very significant simplifications due to the identities \eqref{id1}, \eqref{id2} below. 

Thus both the Feynman diagram and the worldline calculation in this gauge yielded integral representations for diagrams A and B that are manifestly finite
term by term. However those representations are quite different: the one from Feynman diagrams is much more compact, while the one from the
worldline formalism has the advantage of combining the contributions of the planar and the non-planar diagram, something that would be hard to achieve
in the diagrammatic approach. Thus we have chosen to present here both representations, for whatever their worth may be. 
Since our specific purpose will require a fairly high-order computation of the weak-field expansion coefficients of this Lagrangian, we further show how these coefficients
can be obtained from the Lagrangian in an efficient way. Here we use the representation from the diagrammatic approach. 
As expected on general grounds, the coefficients are of the form $r_1 + r_2 \zeta_3$ with rational
numbers $r_1,r_2$, where the $\zeta_3$ comes from the non-planar diagram only. 
We compute the first two coefficients analytically, and four more by numerical integration.  

The organisation of this paper is as follows. In section \ref{sec:ABworldline} we calculate the three-loop quenched EHL in the worldline formalism,
including also a recalculation of the two-loop EHL. The two-loop EHL had been obtained already in \cite{85} using the Feynman diagram approach, but we 
include the worldline calculation here because it displays some interesting simplifications. 
In section \ref{sec:ABfeynman} we calculate the three-loop quenched EHL again in the diagrammatic approach. 
Based on the resulting four-parameter integral representation, we present algorithms for the computation of the weak-field expansion
coefficients in section \ref{sec:compint}. Here for the non-planar contribution we make essential use of the high symmetry of
diagram B, in two ways: first to develop a certain integration-by-parts procedure, and second to rewrite the integrand in Schwinger parameter space
in terms of polynomial invariants of the dihedral group $D_4$, which helps greatly to keep the expressions manageable. 
In section \ref{sec:loworder} we give the results of a calculation of the first six coefficients for both diagrams. Here for the planar diagram A we have analytic results for
all six coefficients, while for the incomparably more difficult non-planar diagram B we have computed the first two coefficients analytically, the remaining four numerically. 
Section \ref{sec:conclusions} gives our summary and outlook. 
There are four appendices: appendix \ref{appconv} gives our conventions. In appendix \ref{appB} we list the momentum integrals appearing in the three-loop
worldline calculation. Appendix \ref{appintegrals} contains the corresponding list for the Feynman diagram calculation. Finally, in appendix \ref{app-invariants} 
we present some elements of the invariant theory of the dihedral group $D_4$, and sketch the derivation of the basis of invariants used in the weak-field expansion
algorithm of diagram B in section \ref{sec:compint}.

\section{Worldline calculation of the quenched three-loop EHL}
\label{sec:ABworldline}
\renewcommand{\theequation}{2.\arabic{equation}}
\setcounter{equation}{0}

\subsection{Worldline representation of 2-photon and 4-photon amplitudes in a constant field}
\label{ehwl}

The starting point for our calculation of the two-loop and three-loop Euler-Heisenberg Lagrangians 
in the worldline formalism
are the following representations of the 2-photon and 4-photon amplitudes in a constant field
(see \cite{41,91} for context and derivation of these formulas).
Define the field strength tensor by $F= ifE$, where

\bear
E &=& \left(\begin{array}{cc}0 & -i \\ i & 0 \end{array}\right)
\label{defE}
\ear
and further $Z\equiv efT$.
Introduce the vacuum worldline Green's functions

\bear
\dot G_{Bij} &=& {\rm sign}(\tau_i-\tau_j)-2 (\tau_i-\tau_j)/T \non\\
G_{Fij} &=& {\rm sign}(\tau_i-\tau_j) \non\\
\label{Gvacuum}
\ear
and the generalized (constant field) Green's functions 

\bear
\bar{\cal G}_{Bij} &=&  
-\frac{1}{2ef}\Bigl(\bar A_{Bij} \Eins +(S_{Bij}-\dot G_{Bij})E\Bigr)
 \non\\
\hat{\dot{{\cal G}}}_{Bij} &=& 
S_{Bij}\Eins +\hat A_{Bij}E 
\nonumber\\
{\cal G}_{Fij} &=& S_{Fij}\Eins + A_{Fij}E
 \non\\
 \label{calGBGF}
\ear
with the scalar, dimensionless  coefficient functions

\bear
S_{Bij}&=&
\frac{\sinh(Z\dot G_{Bij})}{\sinh(Z)} 
\non\\
\bar A_{Bij} &=&
\frac{\cosh(Z \dot G_{Bij})-\cosh{Z}} {\sinh(Z)}
\non\\
\hat A_{Bij} &=& \bar A_{Bij} +\tanh(Z) = \frac{\cosh (Z\dot G_{Bij})}{\sinh(Z)} - \frac{1}{\sinh(Z)\cosh(Z)}
\non\\
S_{Fij} &=&G_{Fij}\frac{\cosh(Z\dot G_{Bij})}{\cosh(Z)}
\non\\
A_{Fij} &=&
G_{Fij}\frac{\sinh(Z\dot G_{Bij})}{\cosh(Z)}
\non\\
\label{defAB}
\ear\no
We note the symmetry properties

\bear
\bar{\cal G}_{Bji} &=& \bar{\cal G}_{Bij}^T, \quad
\hat{\dot{{\cal G}}}_{Bji} = - \hat{\dot{{\cal G}}}_{Bij}^T, \quad
%\ddot{\cal G}_{Bji} = \ddot{\cal G}_{Bji}^T, \quad
{\cal G}_{Fji} = - {\cal G}_{Fij}^T \, .\non\\
\label{symmGBGF}
\ear
%Note also that of the four functions (\ref{defAB}) $A_{B,F}$ have nonvanishing coincidence limits:
%
%\bear
%A_{Bii} &=& \coth(Z)-\frac{1}{Z} \non\\
%A_{Fii} &=& \tanh(Z) \non\\
%\label{coinc}
%\ear
Denote further by

\bear
F_i^{\mu_i \nu_i} &\equiv& k_i^{\mu_i}\varepsilon_i^{\nu_i} - \varepsilon_i^{\mu_i}k_i^{\nu_i}
\label{defFi}
\ear 
the field strength tensor associated to photon $i$, and 
define  the ``super-bicycle of length $n$'' $\dot {\cal G}_S(i_1i_2\cdots i_n)$, involving
a subset $i_1,i_2,\ldots,i_n$ of the integration variables,
by

\bear
\dot {\cal G}_S(i_1i_2) &\equiv& \half {\rm tr} \Bigl(F_{i_1}\hat{\dot{{\cal G}}}_{Bi_1i_2}
F_{i_2}
\hat{\dot{{\cal G}}}_{Bi_2i_1}\Bigr)
 -
 \half {\rm tr} \Bigl(F_{i_1}{\cal G}_{Fi_1i_2}
F_{i_2}
{\cal G}_{Fi_2i_1}\Bigr)
\nonumber\\
\dot {\cal G}_S(i_1i_2\cdots i_n) &\equiv& 
 {\rm tr} \Bigl(F_{i_1}\hat{\dot{{\cal G}}}_{Bi_1i_2}
F_{i_2}
\hat{\dot{{\cal G}}}_{Bi_2i_3}
\cdots
F_{i_n}
\hat{\dot{{\cal G}}}_{Bi_ni_1}
\Bigr) 
\nonumber\\
&&
- 
 {\rm tr} \Bigl(F_{i_1}{\cal G}_{Fi_1i_2}
F_{i_2}
{\cal G}_{Fi_2i_3}
\cdots
F_{i_n}
{\cal G}_{Fi_ni_1}
\Bigr) 
\qquad (n\geq 3)
\nonumber\\
\label{superbicycle}
\ear
Then, the two-photon amplitude in the constant field can be
written as

\bear
\Gamma
[k_1,\varepsilon_1;k_2,\varepsilon_2]
&=&
-e^2
{(2\pi )}^D\delta^D (k_1+k_2)
{ \int_{0}^{\infty}}{dT\over T}
{(4\pi T)}^{-{D\over 2}}
{\e}^{-m^2T}
\nonumber\\
&&\times 
\frac{Z}{\tanh Z}
\int_0^T d\tau_1 d\tau_2
\,\dot {\cal G}_S(12) 
\,\e^{k_1 \bar{\cal G}_{B12}  k_2}
\nonumber\\
\label{rep2phot}
\ear
where $D$ is the space-time dimension. We will set $D=2$ from the beginning, since
dimensional regularization will not become necessary in our calculation.

Further, let us define the ``one-tail'' ${\cal T}(i)$ and ``two-tail'' 
${\cal T}(ij)$ by

\bear
{\cal T}(i) &\equiv& {\sum_{a=1,\ldots,4, a\ne i}}\,\varepsilon_i \hat{\dot {{\cal G}}}_{Bia} k_a \non\\
{\cal T}(ij) &\equiv& 
\sum_{{a\ne i,b\ne j}\atop{(a,b)\ne (j,i)}}
\varepsilon_i \Gphat ia k_a 
\,
\varepsilon_j \Gphat jb k_b 
\nonumber\\&&
+\half
\varepsilon_i \Gphat ij  \varepsilon_j 
\sum_{c\ne i,j}\Bigl( k_i \Gphat ic k_c- k_j \Gphat jc   k_c\Bigr)
\non\\
\label{tails}
\ear
($i,j=1,\ldots ,4$, $i\ne j$).

The four-photon amplitude in the constant field can then be written as \cite{91}
(omitting the global factor $(2\pi)^2\delta(k_1+k_2+k_3+k_4)$)

\bear
&&\Gamma
[k_1,\varepsilon_1;\ldots;k_4,\varepsilon_4]
=
- e^4
{ \int_{0}^{\infty}}{dT\over T}
\frac{{\e}^{-m^2T}}{(4\pi T)}
\frac{Z}{\tanh Z}
\int_0^T 
d\tau_1 d\tau_2 d\tau_3 d\tau_4
\nonumber\\
&&\hspace{20pt}\times
\Bigl(Q^4 + Q^3 + Q^2 +  Q^{22}\Bigr)
\,\exp\biggl\lbrace \half\sum_{i,j=1}^4
k_i \bar{\cal G}_{Bij}  k_j
\biggr\rbrace\nonumber\\
\label{rep4phot}
\ear
where  the upper index on a $Q^i$ denotes the ``cycle content'':

\bear
Q^4 &=& \dot {\cal G}_S(1234) + \dot {\cal G}_S(1243) + \dot {\cal G}_S(1324)
\nonumber\\
Q^3 &=& \dot {\cal G}_S(123){\cal T}(4) + \dot {\cal G}_S(234){\cal T}(1) + \dot {\cal G}_S(341){\cal T}(2) +\dot {\cal G}_S(412){\cal T}(3)  
\nonumber\\
Q^2 &=& \dot {\cal G}_S(12){\cal T}(34)+ 
\dot {\cal G}_S(13){\cal T}(24)+ 
\dot {\cal G}_S(14){\cal T}(23)
\nonumber\\&&
+\dot {\cal G}_S(23){\cal T}(14)+ 
\dot {\cal G}_S(24){\cal T}(13)+ 
\dot {\cal G}_S(34){\cal T}(12)
\nonumber\\
Q^{22} &=&
 \dot {\cal G}_S(12)\dot{\cal G}_S(34)+ 
\dot {\cal G}_S(13)\dot{\cal G}_S(24)+ 
\dot {\cal G}_S(14)\dot{\cal G}_S(23)
\nonumber\\
\label{defQ}
\ear
Let us mention that each of the sixteen terms appearing in the
decomposition (\ref{defQ}) gives a contribution to the four photon amplitude that
is separately gauge invariant \cite{26,41,91}. 

The representations (\ref{rep2phot}), (\ref{rep4phot}) hold
{\sl mutatis mutandis} also in 4D QED, however in the 2D case
they are particularly useful,
because here all matrices appearing in the above expressions (the $F_i$'s and all worldline
Green's functions) commute with each other. This is because in two dimensions all antisymmetric matrices are multiples
of each other, and the Green's functions involve only the matrix $F$. 
Thus any matrix, or product of matrices, appearing in our calculations below is of the form 
$M=a\Eins + bE$, which also entails the useful identities

\bear
M + M^T &=& \Eins {\rm tr}M \, ,
\label{idMMt}
\\
k_iMk_i &=& \half k_i^2 {\rm tr}M  \, .
\label{idkMk}
\ear
These simple observations will lead to many simplifications in the following calculations.
In particular,  the commutativity implies that, for even $n$, the super-bicycles (\ref{superbicycle}) factorize as

\bear
\dot {\cal G}_S(i_1i_2) &=& \fourth {\rm tr} \Bigl(\hat{\dot{{\cal G}}}_{Bi_1i_2}
\hat{\dot{{\cal G}}}_{Bi_2i_1}-{\cal G}_{Fi_1i_2}{\cal G}_{Fi_2i_1}\Bigr)
{\rm tr}\Bigl(F_{i_1}F_{i_2}\Bigr)
\nonumber\\
\dot {\cal G}_S(i_1i_2\cdots i_n) &=& 
\half {\rm tr}
\Bigl(
\hat{\dot{{\cal G}}}_{Bi_1i_2}
\hat{\dot{{\cal G}}}_{Bi_2i_3}
\cdots
\hat{\dot{{\cal G}}}_{Bi_ni_1}
- {\cal G}_{Fi_1i_2}
{\cal G}_{Fi_2i_3}
\cdots
{\cal G}_{Fi_ni_1}
\Bigr) 
\nonumber\\&&\times
{\rm tr}\Bigl(F_{i_1}F_{i_2}\cdots F_{i_n}\Bigr)
\qquad \qquad (n\geq 4)
\nonumber\\
\label{superbicyclefactored}
\ear

The representations (\ref{rep2phot}), (\ref{rep4phot}) hold off-shell, so that the one-loop
amplitudes can be used to construct (quenched) higher-loop photon amplitudes by sewing
off pairs of photons, say, the photon legs with index $i$ and $j$: Using an arbitrary
covariant gauge with gauge parameter $\xi$,
this is implemented by setting

\bear
k_j &\to& -k_i, \quad \varepsilon_i^{\mu}\varepsilon_j^{\nu} \to  
\delta^{\mu\nu} - (1-\xi)\frac{k_i^{\mu}k_i^{\nu}}{k_i^2}
\label{sewing}
\ear
and adding the integration $\int d^2 k_i/(4\pi^2 k_i^2)$ (there is also a combinatorial factor of $\half$
for each pair of legs).

Thus in the following we will construct the two-loop Euler-Heisenberg Lagrangian from
the one-loop vacuum polarization tensor, and the three-loop EHL from 
the one-loop four-photon amplitude. In the worldline formalism, one could construct these
higher loop Euler-Heisenberg Lagrangians also directly using the concept of multi-loop
worldline Green's functions \cite{8,15,41}; however, this would obscure the 
decomposition (\ref{defQ}), which we will find very useful in the following. 
This is because, first, it will allow us to substantially reduce the number of
terms in the integrand using the symmetries of the problem; and second,
because the gauge invariance term-by-term of this decomposition implies that for each term
we can use a different gauge parameter $\xi$ in the sewing procedure
(\ref{sewing}).

\subsection{One-loop vacuum polarization tensor in a constant field}
\label{vp}

Let us start with calculating the vacuum polarization tensor,
related to the two-photon amplitude (\ref{rep2phot}) by

\bear
\Gamma
[k_1,\varepsilon_1;k_2,\varepsilon_2] &=& (2\pi)^2\delta(k_1+k_2)\varepsilon_1 \Pi\varepsilon_2 \, .
\label{defPi}
\ear
We can simplify this calculation by observing that the constant-field vacuum polarization
tensor carries  the transversal projector familiar from the vacuum case,

\bear
\Pi_\mn (k) &=& \Bigl(\delta_{\mn}k^2 - k_\mu k_\nu \Bigr)\Pi(k) \, .
\label{reducePi}\ear
We note that this is another peculiarity of the 2D case, and due to the fact that no other symmetric and transversal tensor of second degree can be built from $\delta_{\mu \nu},k_{\mu}$,
and $F_{\mn}$. In 4D QED the tensor $\Pi_{\mu\nu}$ in a constant field 
involves several independent transversal structures  (see, e.g., \cite{ditgie-book}). 

\no
Thus $\Pi_{\mu}^{\mu}(k) = k^2\Pi(k)$, and it is
sufficient to calculate the trace of $\Pi_\mn$, corresponding to the replacement
$\varepsilon_1^{\mu}\varepsilon_2^{\nu}\to \delta^{\mn}$.
Making this replacement in $F_{1,2}$, setting $k=k_1=-k_2$, 
and using (\ref{superbicycle}), we are led to compute

\bear
\half {\rm tr}(F_1F_2) &=& k^2
\label{F1F2}
\ear
and

\bear
\half {\rm tr} \Bigl(\hat{\dot{{\cal G}}}_{B12}\hat{\dot{{\cal G}}}_{B21}
-
{\cal G}_{F12}{\cal G}_{F21}\Bigr)
&=&
\hat A_{B12}^2 - S_{B12}^2  - (A_{F12}^2 - S_{F12}^2) 
\nonumber\\
&=&
2 \frac{\cosh Z - \cosh(Z\dot G_{B12})}{\cosh Z \sinh^2 Z} \,.
\label{2loopGS}
\ear
In the exponent we get

\bear
k_1 \bar {\cal G}_{B12}  k_2 =  \bar A_{B12}\frac{k^2}{2ef}
= -  \frac{\cosh Z - \cosh(Z\dot G_{B12})}{\sinh Z} \frac{k^2}{2ef} \, .
\label{2loopexponent}
\ear
Putting things together,

\bear
\Pi_\mn (k) &=& -\Bigl(\delta_{\mn}k^2 - k_\mu k_\nu \Bigr)\frac{e^2}{2\pi}
\int_0^{\infty}\frac{dT}{T^2}\e^{-m^2T} \frac{Z}{\sinh^3 Z} 
\nonumber\\
&& \times
\int_0^Td\tau_1 d\tau_2 \Bigl(\cosh Z - \cosh(Z\dot G_{B12})\Bigr)
\,\e^{ -  \frac{\cosh Z - \cosh(Z\dot G_{B12})}{\sinh Z} \frac{k^2}{2ef}} \, .
\nonumber\\
\label{vpwlfinal}
\ear
Finally we use, as usual in this type of calculations \cite{41},  the translation invariance of the worldline Green's functions 
to put $\tau_2=0,\tau_1=\tau$, and rescale $\tau = Tu$.
With a further change of variables from $T$ to $Z$ 
and from $m^2$ to $\kappa = m^2/(2ef)$ we obtain our final form for the vacuum polarization tensor,

\bear
\Pi_{\mu \nu}(k) &=& - { e \over 2 \pi f} (\delta_{\mu \nu}k^2-k_{\mu} k_{\nu}) 
{ \int_{0}^{\infty}}{dZ}
\e^{-2\kappa Z}
\frac{Z}{\sinh^3 Z}
\int_0^1
du
\non\\
&&\hspace{-30pt}
\times 
\,{\rm exp}\biggl\lbrack - \frac{k^2}{ef} {{\cosh Z - \cosh(\dot G_{B12}Z)}\over{2\sinh Z }}\biggr\rbrack
\big[\cosh Z -  \cosh ( \dot{G}_{B12}Z)\big]
\nonumber\\
\label{Pispinfin}
\ear
where now $\dot G_{B12} = 1- 2 u$.

\subsection{Two-loop EH Lagrangian}
\label{wl2loop}

We now construct the two-loop EH Lagrangian from the polarization tensor.
Using (\ref{Pispinfin}) in the sewing procedure gives
(note that $\cosh Z \geq  \cosh ( \dot{G}_{B12}Z)$)

\bear
{\cal L}^{(2)}(f)
&=&
\half \int \frac{d^2k}{(2\pi)^2k^2}
\Bigl(\delta^{\mu\nu} - (1-\xi)\frac{k^{\mu}k^{\nu}}{k^2}\Bigr)
\Pi_{\mu \nu}(k) 
\non\\
&=& -{e\over 4 \pi f} 
\int \frac{d^2k}{(2\pi)^2}
{\int_{0}^{\infty}}{dZ}
\e^{-2\kappa Z}
\frac{Z}{\sinh^3 Z}
\int_0^1
du
\non\\
&&\hspace{-30pt}
\times 
\,{\rm exp}\biggl\lbrack - \frac{k^2}{ef} {{\cosh Z - \cosh(\dot G_{B12}Z)}\over{2\sinh Z }}\biggr\rbrack
\big[\cosh Z -  \cosh ( \dot{G}_{B12}Z)\big]
\, .
\nonumber\\
\label{Lspin2wl1}
\ear
Remarkably, after the gaussian $k$ - integration all $u$ - dependence has already cancelled out, and one is left with a 
one-parameter-integral:

\bear
{\cal L}^{(2)}(f)
&=& -{e^2\over 8\pi^2}
{\int_{0}^{\infty}}{dZ}
\e^{-2\kappa Z}
\frac{Z}{\sinh^2 Z}
\, .
\nonumber\\
\label{Lspin2wl2}
\ear
We note that the lowest order ({\it f} -- independent = vacuum) term has an UV divergence
at $Z=0$. Subtracting this lowest order term, that is replacing $1/\sinh^2 Z$ by
$1/\sinh^2 Z - 1/Z^2$, we can apply the standard integral formula \cite{51} 

\bear
I(\kappa,\epsilon) &\equiv& \int_0^{\infty}dx\,\e^{-2\kappa x}
\Bigl\lbrack \frac{1}{\sinh^{2+\epsilon}x} -\frac{1}{x^{2+\epsilon}} \Bigr\rbrack  
\nonumber\\
&=&
- 2 \xi(\kappa) + \Bigl\lbrack \frac{\xi^2(\kappa)}{\kappa} + 2\bigl(1-\gamma - \ln (2\kappa)\bigr)\xi(\kappa) - \frac{1}{4\kappa}\Bigr\rbrack \epsilon
+ O(\epsilon^2)
\, .
\nonumber\\
\label{Ikeps}
\ear
Thus we find

\bear
{\cal L}^{(2)}(f)
&=& -{e^2\over 8\pi^2}(-\half \partder{}{\kappa})I[\kappa,0] = -{e^2\over 8\pi^2}\xi'(\kappa)
\, .
%= {e^2\over 8\pi^2}(\tilde \psi(\kappa) + \kappa \tilde\psi'(\kappa))
\nonumber\\
\label{L2spinwlfinrep1}
\ear
This agrees, up to the irrelevant divergent vacuum term, with the result of the Feynman diagram calculation in \cite{81} that we quoted already
in the introduction, eq. (\ref{L2spinwlfin}).

\subsection{Three-loop quenched EH Lagrangian}
\label{3loopquenched}

We proceed to the construction of the quenched three-loop EH Lagrangians, starting from the
worldline representation (\ref{rep4phot}) of the one-loop four-photon amplitudes in the field.  
We apply  the sewing procedure
(\ref{sewing}) with $k_1=-k_2 =: k$ and $k_3=-k_4=:l$. 
The universal exponential factor in (\ref{rep4phot}) can then be written as

\bear
\half \sum_{i,j=1}^4k_i \bar {\cal G}_{Bij}  k_j
&=&
- \half (k,l) {\cal M} (k,l)
\label{expfact}
\ear
where 

\bear
{\cal M} &=& \left(\begin{array}{cc} {\cal A} & - {\cal C} \\ -{\cal C}^T & {\cal B}\end{array}\right)
\label{defM}
\ear
with

\bear
{\cal A} &=&  \bar {\cal G}_{B12} +  \bar {\cal G}_{B21} -  \bar {\cal G}_{B11}  -  \bar {\cal G}_{B22} = \alpha\Eins\non\\
{\cal B} &=&  \bar {\cal G}_{B34} +  \bar {\cal G}_{B43} -  \bar {\cal G}_{B33}  -  \bar {\cal G}_{B44} = \beta\Eins \non\\
{\cal C} &=& \bar {\cal G}_{B13} +  \bar {\cal G}_{B24} -  \bar {\cal G}_{B14}  -  \bar {\cal G}_{B23} = \gamma\Eins -  \tilde\gamma E \non\\
\label{defABC}
\ear 
and

\bear
\alpha &=& T\,{{\cosh Z - \cosh(\dot G_{B12}Z)}\over{Z\sinh Z }} \nonumber\\
\beta &=& T\,{{\cosh Z - \cosh(\dot G_{B34}Z)}\over{Z\sinh Z }} \nonumber\\
\gamma &=& -\frac{T}{2Z\sinh Z}
\Bigl\lbrack\cosh(\dot G_{B13}Z) + \cosh(\dot G_{B24}Z) -\cosh(\dot G_{B14}Z) - \cosh(\dot G_{B23}Z) \Bigr\rbrack\nonumber\\
\tilde\gamma &=& \frac{T}{2Z\sinh Z}
\Bigl\lbrack
\sinh(\dot G_{B13}Z) + \sinh(\dot G_{B24}Z) -\sinh(\dot G_{B14}Z) - \sinh(\dot G_{B23}Z) \nonumber\\&& \hspace{60pt}
-  \bigl( \dot G_{B13} +  \dot G_{B24} - \dot G_{B14}  -  \dot G_{B23}\bigr) \sinh Z
\Bigr\rbrack 
\label{defabcd}
\ear
We will also need the determinant and the inverse of $\cal M$,

\bear
{\rm det}{\cal M} &=& \Delta^2, \quad
\Delta \equiv \alpha\beta - \gamma^2 + \tilde\gamma^2, \nonumber\\
\label{detM}
\ear

\bear
{\cal M}^{-1} &=& \frac{1}{\Delta}\left(\begin{array}{cc} {\cal B} &  {\cal C} \\ {\cal C}^T & {\cal A}\end{array}\right)
= 
 \frac{1}{\Delta}
\left(\begin{array}{cc} \beta \Eins & \gamma\Eins - \tilde\gamma E \\   \gamma\Eins + \tilde\gamma E& \alpha\Eins \end{array}\right).
\label{Minv}
\ear
In the following we will often treat as numbers matrices which are proportional
to the unit matrix, such as $\hat {\dot {{\cal G}}}_{B12} \hat{\dot {{\cal G}}}_{B21}$. 

Further, we observe that, after the sewing, the resulting total worldline integral is invariant under the exchanges
$\tau_1 \leftrightarrow \tau_2$,  $\tau_3 \leftrightarrow \tau_4$, and $(\tau_1,\tau_2)\leftrightarrow (\tau_3,\tau_4)$.
This implies that, of the sixteen terms appearing in the decomposition (\ref{defQ}), only seven are independent and
need to be computed; namely, we can now make the replacements

\bear
Q^4 &\to& 2\dot {\cal G}_S(1234) +  \dot {\cal G}_S(1324)
\nonumber\\
Q^3 &\to & 4\dot {\cal G}_S(123){\cal T}(4) \nonumber\\
Q^2 &\to & 2\dot {\cal G}_S(12){\cal T}(34)+ 
4\dot {\cal G}_S(13){\cal T}(24)
\nonumber\\
Q^{22} &\to&
 \dot {\cal G}_S(12)\dot{\cal G}_S(34)+ 
2\dot {\cal G}_S(13)\dot{\cal G}_S(24)
\nonumber\\
\label{Q3loop}
\ear
We will now first work out the effect of the sewing replacements

\bear
&& k_1 \to k, k_2 \to -k, k_3\to l, k_4\to -l, \varepsilon_1^{\mu}\varepsilon_2^{\nu}\to \delta^{\mu\nu}- (1-\xi)\frac{k^{\mu}k^{\nu}}{k^2},\quad
\varepsilon_3^{\mu}\varepsilon_4^{\nu}\to \delta^{\mu\nu}- (1-\xi) \frac{l^{\mu}l^{\nu}}{l^2}
\nonumber\\
\label{sewreplace}
\ear

\no
on these seven structures. This is simplest for the ones in $Q^4$ and $Q^{22}$; e.g.,
in $\dot {\cal G}_S(1324)$ we can use the commutativity of all matrices to write

\bear
 {\rm tr} \Bigl(F_1
\hat{\dot{{\cal G}}}_{B13}
F_3
\hat{\dot{{\cal G}}}_{B32}
F_2
\hat{\dot{{\cal G}}}_{B24}
F_4
\hat{\dot{{\cal G}}}_{B41}
\Bigr) 
&=&
\fourth {\rm tr}(F_1F_2){\rm tr}(F_3F_4) {\rm tr} 
\Bigl(
\hat{\dot{{\cal G}}}_{B13}
\hat{\dot{{\cal G}}}_{B32}
\hat{\dot{{\cal G}}}_{B24}
\hat{\dot{{\cal G}}}_{B41}
\Bigr)
\nonumber\\
&=&
k^2l^2 {\rm tr}\Bigl(
\hat{\dot{{\cal G}}}_{B13}
\hat{\dot{{\cal G}}}_{B32}
\hat{\dot{{\cal G}}}_{B24}
\hat{\dot{{\cal G}}}_{B41}
\Bigr)
\, .
\label{simptrace}
\ear
In this way, we obtain (independently of $\xi$),

\begin{eqnarray}
\dot {\cal G}_S(1234) &=& k^2l^2 {\rm tr}\Bigl(
\lbrace{1243}\rbrace_S
\Bigr)
\nonumber\\
\dot {\cal G}_S(1324) &=& 
k^2l^2 {\rm tr}\Bigl(
\lbrace{1324}\rbrace_S
\Bigr)
\nonumber\\
\dot {\cal G}_S(12)\dot{\cal G}_S(34) &=&\fourth k^2l^2
{\rm tr}\Bigl(
\lbrace 12\rbrace_S
\Bigr)
{\rm tr}\Bigl(
\lbrace 34 \rbrace_S
\Bigr)
\nonumber\\
\dot {\cal G}_S(13)\dot{\cal G}_S(24) &=&
\fourth k^2l^2
{\rm tr}\Bigl(
\lbrace 13\rbrace_S
\Bigr)
{\rm tr}\Bigl(
\lbrace 24 \rbrace_S
\Bigr)
\nonumber\\
\label{Q4Q22}
\end{eqnarray}
where we have further introduced the abbreviation

\bear
\lbrace i_1i_2\ldots i_n \rbrace_S &\equiv&
\Gphat{i_1}{i_2}\Gphat{i_2}{i_3}\cdots\Gphat{i_n}{i_1} 
-
{\cal G}_{Fi_1i_2}{\cal G}_{Fi_2i_3} \cdots {\cal G}_{Fi_ni_1}
\, .
\nonumber\\
\label{defbrace}
\ear
Proceeding to $Q^3$, here we also use $E^2 = \Eins$ to rewrite

\bear
{\rm tr} \Bigl(F_1
\hat{\dot{{\cal G}}}_{B12}
F_2
\hat{\dot{{\cal G}}}_{B23}
F_3
\hat{\dot{{\cal G}}}_{B31}
\Bigr) 
&=&
{\rm tr} \Bigl(E^2 F_1
\hat{\dot{{\cal G}}}_{B12}
F_2
\hat{\dot{{\cal G}}}_{B23}
F_3
\hat{\dot{{\cal G}}}_{B31}
\Bigr) 
\nonumber\\
&=&\fourth
{\rm tr} \bigl(F_1F_2\bigr)
{\rm tr} \bigl(F_3E\bigr)
{\rm tr}
\Bigl(E
\hat{\dot{{\cal G}}}_{B12}
\hat{\dot{{\cal G}}}_{B23}
\hat{\dot{{\cal G}}}_{B31}
\Bigr) 
\non\\
&=&
k^2\varepsilon_3 E l
\,
{\rm tr}
\Bigl(E
\hat{\dot{{\cal G}}}_{B12}
\hat{\dot{{\cal G}}}_{B23}
\hat{\dot{{\cal G}}}_{B31}
\Bigr) 
\, .
\label{Q3step1}
\ear
After the sewing of $\varepsilon_3$ with $\varepsilon_4$ this gives

\bear
\dot {\cal G}_S(123){\cal T}(4) &=&
k^2
{\rm tr}
\Bigl(E
\lbrace 123 \rbrace_S
\Bigr) 
\Bigl(
l E  (\hat{\dot{{\cal G}}}_{B42}-\hat{\dot{{\cal G}}}_{B41})  k
- l E \hat{\dot{{\cal G}}}_{B43} l
\Bigr)
\non\\
\label{Q3step2}
\ear
(the term involving $\xi$ drops out here because it contains a factor $lEl=0$).

Coming to the terms in $Q^2$, here things are more delicate. 
It turns out that, to avoid the
appearance of spurious infrared divergences in the $k,l$
integrations, one should use the gauge $\xi = -1$.
This leads to

\bear
\dot {\cal G}_S(12){\cal T}(34) &=&
\half k^2 {\rm tr}\Bigl(\lbrace 12\rbrace_S\Bigr)\Bigl[
 k (\Gphat23-\Gphat13)P_{34} (\Gphat41 -\Gphat42 ) k      \non\\
&& + l  \Bigl( \Gphat43 (\Gphat41 - \Gphat42) - \Gphat34 (\Gphat31 - \Gphat32) \non \\
&& + ({\rm tr}\Gphat34)(\Gphat41 - \Gphat42  + \Gphat31 - \Gphat32) \Bigr)  k  \Bigr]
\label{Q212}
\ear
where

\bear
P_{34}^{\mu\nu} &=& \delta^{\mu\nu} - 2 \frac{l^{\mu}l^{\nu}}{l^2} \nonumber\\
\label{P34}
\ear
and

\bear
\dot {\cal G}_S(13){\cal T}(24) &=&
\half
{\rm tr}\Bigl(\lbrace 13\rbrace_S\Bigr)
\biggl[
\Bigl(l\Gphat21 k+l(\Gphat23 -\Gphat24)l\Bigr)
\Bigl(k(\Gphat41 -\Gphat42 )k +k\Gphat43 l\Bigr)
 \non\\
 &&
 + k l \Bigl(k\Gphat 12 + l(\Gphat32 -\Gphat42)\Bigr)
 \Bigl( (\Gphat 41 - \Gphat 42)k +\Gphat43 l\Bigr)
 \non\\
 &&
 - l\Gphat24 l k\Gphat42 k -k l \, l \Gphat42\Gphat42 k
 \non\\
 &&
 +\half \Bigl(l\Gphat24 k - k l {\rm tr} (\Gphat24) \Bigr)
 \Bigl(-k\Gphat21 k -k\Gphat23 l +l\Gphat41 k + l\Gphat43 l \Bigr)
\biggr]
\, .
\nonumber\\
\label{Q213}
\ear
The momentum integrals can now be easily performed, without encountering any IR divergences,
making them gaussian by exponentiating each denominator, 

\bear
\frac{1}{k^2} = \int_0^{\infty} dx \,\e^{-x k^2}, \qquad \frac{1}{l^2} = \int_0^{\infty} dy \,\e^{-y l^2}
\, .
\label{exponentiate}
\ear
The parameter integrals over $x,y$ are convergent since $\alpha,\beta,\Delta >0$.
A list of all integrals is given in appendix \ref{appB}.

The exception is the first term in the square brackets in (\ref{Q212}) 
which contains an IR divergence in the $l$ integral for any gauge except for our choice $\xi =-1$;
here care must be taken, and we therefore present its calculation in some detail.
We abbreviate $A =  \Gphat23-\Gphat13$ and $B=\Gphat41 -\Gphat42$. The $1/k^2$ from the
propagator here cancels against the $k^2$ in the prefactor of (\ref{Q212}). Thus the $k$ integral is
finite, and performing it first gives

\bear
\int \frac{d^2k}{(2\pi)^2} kAP_{34}Bk\,\e^{-\half \alpha k^2 + k{\cal C}l}
=
\frac{1}{2\pi\alpha} \Bigl(\frac{1}{\alpha}{\rm tr}(AP_{34}B) + \frac{1}{\alpha^2}lC^TAP_{34}BCl\Bigr)
 \,\e^{\frac{1}{2\alpha}({\cal C}l)^2}
\, . 
\non\\
\label{kint}
\ear
The first term in the brackets, which is the IR-critical one, vanishes since ${\rm tr}(AP_{34}B)={\rm tr}(P_{34}BA)$,
$BA$ is a linear combination of $\Eins$ and $E$, and ${\rm tr}P_{34}={\rm tr}(P_{34}E) =0$.
In the second term, we can use the identity (\ref{idkMk}) to write

\bear
lC^TAP_{34}BCl &=& lC^TABCl - \frac{2}{l^2}(lC^TAl)(lBCl) 
\non\\
&=& \half l^2 \bigl\lbrack\tr(C^TABC) - \tr(C^TA)\tr(BC)\bigr\rbrack
\, .
\label{rewritesecond}
\ear
Thus also the $l$ - propagator cancels, and doing the $l$ - integral we obtain for the total integral the result

\bear
\int \frac{d^2k}{(2\pi)^2}\int\frac{d^2l}{(2\pi)^2} kAP_{34}Bk\frac{1}{l^2}\,\e^{-\half \alpha k^2 - \half\beta l^2 + k{\cal C}l}
=
\frac{\tr(C^TABC) - \tr(C^TA)\tr(BC)}{8\pi^2\alpha^2\Delta}
\, .
\non\\
\label{intcritfin}
\ear
All the other integrals are straightforward.
In this way we arrive at our final result for the quenched part of the three-loop Lagrangian, corresponding to the sum of the diagrams $A$ and $B$ of Fig. \ref{fig-ABC}:

\bear
{\cal L}^{3(A+B)}(f)
&=&
-\frac{e^4}{(4\pi)^3}
{ \int_{0}^{\infty}}{dT\over T^2}
{\e}^{-m^2T}
\frac{Z}{\tanh Z}
\prod_{i=1}^4 \int_0^T 
d\tau_i
\nonumber\\&&\times
\Bigl(2I_{1234}+I_{1324}+4I_{123}+2I_{12}+4I_{13}+I_{12,34}+2I_{13,24}\Bigr)
\non\\
\label{L3quenchedscal}
\ear
where

\bear
I_{ijkl} &=& \frac{{\rm tr}\bigl(\lbrace ijkl\rbrace_S\bigr)}{\Delta} \nonumber\\
I_{ij,kl} &=& \frac{{\rm tr}\bigl(\lbrace ij \rbrace_S\bigr){\rm tr}\bigl(\lbrace kl \rbrace_S\bigr)}{4\Delta}\nonumber\\
I_{123}&=& -
\frac{
{\rm tr}
\Bigl(E
\lbrace 123 \rbrace_S
\Bigr)}
{2\Delta} 
{\rm tr}\Bigl(E\Gphat43 +\frac{1}{\alpha}E(\Gphat41 - \Gphat42 ){\cal C}\Bigr)
\nonumber\\
I_{12} &=&  
 - \frac{{\rm tr}\bigl(\lbrace 12\rbrace_S\bigr)}{4\alpha^2\Delta}
\Bigl\lbrack {\rm tr}\bigl({\cal C}^T(\Gphat13-\Gphat23)(\Gphat41 - \Gphat42){\cal C}\bigr)
\non\\
&& \qquad\qquad -
{\rm tr}\bigl({\cal C}^T(\Gphat13-\Gphat23)\bigr){\rm tr}\bigl((\Gphat41 - \Gphat42){\cal C}\bigr)
\Bigr\rbrack
\non\\
&& 
 - \frac{{\rm tr}\bigl(\lbrace 12\rbrace_S\bigr)}{4\alpha\Delta}
\Bigl\lbrack {\rm tr}\bigl({\cal C}^T(\Gphat13-\Gphat23)\Gphat43\bigr)
-{\rm tr}\bigl({\cal C}^T(\Gphat13-\Gphat23)\bigr){\rm tr}\bigl(\Gphat43\bigr)
\non\\
&& \qquad\qquad\quad
 -  {\rm tr}\bigl(\Gphat43(\Gphat41-\Gphat42){\cal C}\bigr)
+ {\rm tr}\bigl(\Gphat43\bigr){\rm tr}\Bigl((\Gphat41-\Gphat42){\cal C}\bigr)
\Bigr\rbrack
\non\\
I_{13} &=& 
 \frac{{\rm tr}\bigl(\lbrace 13\rbrace_S\bigr)}{2}
 \Biggl\lbrace
 \frac{1}{4\Delta}
 \Bigl\lbrack
 \tr (\Gphat41 -\Gphat42)\tr(\Gphat23 -\Gphat24)-\tr (\Gphat42)\tr( \Gphat24)
 \Bigr\rbrack
 \non\\&&\qquad\qquad
 + \frac{h_{3,1}}{4} 
 \Bigl\lbrack
 -\tr (\Gphat41 -\Gphat42)\tr(\Gphat12 {\cal C}^T) 
 + 2\gamma \tr \bigl(\Gphat12 (\Gphat41 -\Gphat 42)\bigr)
 \non\\&&\qquad\qquad\qquad
 -\half\tr (\Gphat12)\Bigl(\tr (\Gphat42 {\cal C}^T)-2\gamma\tr (\Gphat 42)\Bigr) 
 \Bigr\rbrack
  \non\\&&\qquad\quad
 + \frac{h_{1,3}}{4} 
 \Bigl\lbrack
 \tr (\Gphat23 -\Gphat24)\tr(\Gphat43 {\cal C}^T) 
 + 2\gamma \tr \bigl(\Gphat43 (\Gphat32 -\Gphat 42)\bigr)
 \non\\&&\qquad\qquad\qquad
 +\half\tr (\Gphat34)\Bigl(\tr (\Gphat42 {\cal C}^T)- 2\gamma\tr (\Gphat 42)\Bigr) 
 \Bigr\rbrack
  \non\\&&\qquad\quad
  + \frac{h_{2,2}^{kk,ll}}{8} \tr \Bigl\lbrack
  \Gphat43(\Gphat21+\Gphat12) +\Gphat32\Gphat41 
  \non\\&&\qquad\qquad\qquad
  -\half\bigl(\Gphat23 +\Gphat 14\bigr)\bigl(3\Gphat24 - \tr(\Gphat24)\Eins \bigr)
  \Bigr\rbrack
  \non\\&&\qquad\quad
+ \frac{h_{2,2}^{kl,kl}}{8}  \Bigl\lbrack
\tr\bigl( (\Gphat32\Gphat41-\Gphat32\Gphat42-\Gphat42\Gphat41){\cal C}^2\bigr)
 \non\\&&\qquad\qquad\qquad
+\half\tr \Bigl( \bigl(\Gphat24 - \tr(\Gphat24)\Eins \bigr)\bigl(\Gphat 32+\Gphat41\bigr){\cal C}^2\Bigr)
 -\tr(\Gphat34 {\cal C})\tr(\Gphat21{\cal C})
  \non\\&&\qquad\qquad\qquad
 + 2\gamma \tr\bigl( (\Gphat32\Gphat41-\Gphat32\Gphat42-\Gphat42\Gphat41+\Gphat21\Gphat34){\cal C}\bigr)
  \non\\&&\qquad\qquad\qquad
  +\half\tr\bigl((\Gphat32+\Gphat41){\cal C}\bigr)\tr\Bigl(\bigl(\Gphat24- \tr(\Gphat24)\Eins \bigr){\cal C}\Bigr) 
\Bigr\rbrack
 \Biggr\rbrace
\label{Ifin}
\ear
where

\bear
h_{3,1} &=& \frac{1}{\alpha\Delta}
\nonumber\\
h_{2,2}^{kk,ll} &=& \frac{1}{\Delta} + \frac{\ln (\alpha\beta/\Delta)}{\gamma^2 -\tilde\gamma^2}
\nonumber\\
h_{2,2}^{kl,kl} &=& 
 \frac{1}{(\gamma^2 -\tilde\gamma^2)\Delta} - \frac{\ln (\alpha\beta/\Delta)}{(\gamma^2 -\tilde\gamma^2)^2}
\nonumber\\
h_{1,3} &=&  \frac{1}{\beta\Delta}
\nonumber\\
\label{defh}
\ear
Note that in the calculation of $I_{13}$ we have used  the identities (\ref{idMMt}), (\ref{idkMk}),
the fact that all matrices are commuting, 
and that ${\hat{{\dot {{\cal G}}}}_{Bij}}  + \hat{{\dot {{\cal G}}}}_{Bij}^T  \sim \Eins$.

Note also that our final integrand is still given in the decomposition
corresponding to (\ref{rep4phot}),(\ref{Q3loop}), with the subscript on a term
denoting its cycle content.

\section{Feynman diagram calculation of the quenched three-loop EHL}
\label{sec:ABfeynman}
\renewcommand{\theequation}{3.\arabic{equation}}
\setcounter{equation}{0}

In this section, we will calculate the quenched part of the three-loop EHL another time, now using the standard formalism. 
In terms of Feynman diagrams, the quenched part is
given by diagrams A and B of figure \ref{fig-ABC}. 

\subsection{Definitions}

We use the 2D electron propagator in a field with constant field strength tensor 
$F=\bigl( \begin{smallmatrix}   0 & f\\-f & 0 \end{smallmatrix} \bigr)$, where
$f>0$. The electron propagator in the field can, using the proper-time representation and Fock-Schwinger gauge,
be written as (see appendix \ref{appconv} for our conventions) 
 
\be \nn
G(p) = \int_{0}^{\infty} dz~ \varphi(z,p) g(z,p)
\label{defG}
\ee
where

\begin{eqnarray}
\varphi(z,p) &\equiv& \frac{1}{ef} \e^{- ( 2 \kappa z + \frac{\tanh z}{ef} p^2) } \frac{1}{\cosh z} \nonumber\\
g(z,p) &\equiv& m \e^{\sigma_3 z} - i \frac{\p}{\cosh z} 
\end{eqnarray}
($\kappa = m^2 / 2ef$).
In the following we will often omit the arguments in $\varphi(z,p)$ and $g(z,p)$ and replace them by
superscripts, e.g., $\hat g \equiv g(\hat z,\hat p)$.
Moreover, we will abbreviate $\gamma \equiv 1/\cosh z, \gamma'\equiv1/\cosh z'$, etc.
Greek indices run over the values 1,2 as usual and we use the shorthand 
$x \wedge y \equiv x_\al y_\bt \ep_{\al \bt}$. 
As in our three-loop calculation in the worldline formalism of the previous section, 
in the Feynman diagram calculation, too, at the three-loop level it turns out to be essential for avoiding spurious IR divergences
to take both photon¤ propagators  in the ``traceless'' gauge $\xi = -1$, where

\bear
D_{\alpha \beta} (k) &=& \frac{1}{k^2}\bigl(\delta_{\alpha \beta} - 2\frac{k_{\alpha} k_{\beta}}{k^2} \bigr). \qquad
\label{props}
\ear
In this gauge one has, in $D=2$ and for any $n\geq 0$, the extremely useful identities

\bear
\sigma_{\alpha}\sigma_{\nu_1}\cdots \sigma_{\nu_{2n}}\sigma_{\beta}D_{\alpha \beta} (k) 
&=& 0 \, , \label{id1} \\
\sigma_{\alpha}\sigma_{\nu_1}\cdots \sigma_{\nu_{2n+1}}\sigma_{\beta}D_{\alpha \beta} (k) 
&=& -2 \frac{1}{k^4} 
\slash k
\sigma_{\nu_1}\cdots \sigma_{\nu_{2n+1}} \slash k \, .
\label{id2}
\ear

\subsection{Diagram A} 

\begin{figure}[h]
\begin{center}
\includegraphics[angle=0, width=0.3\textwidth]{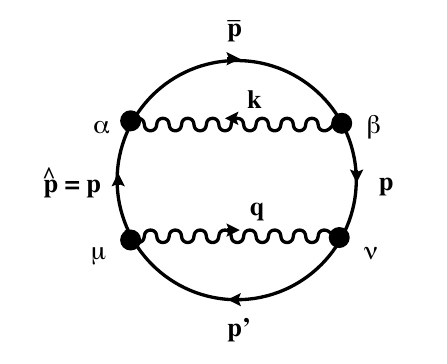}
\end{center}
\caption{Parametrization of diagram A.}
\label{A}
\end{figure}

We start with the simpler one, diagram A. We parametrize this diagram as in figure \ref{A}, 
with independent momenta $p,q,k$, so that the kinematic relations are

\be \label{kin}
p' = p+q, \quad \ph = p, \quad \pb = p+k
\ee

\noindent
The Feynman rules give for this diagram the following expression:

\begin{eqnarray} \label{PsiA}
{\cal L}^{3A} (f) &=&  - 2 \frac{e^4}{4}  \int_{0}^{\infty}dz d\zb d\zh dz' \int \frac{d^2q}{(2\pi)^2} \int  \frac{d^2 p}{(2 \pi)^2}\int \frac{d^2 k}{(2 \pi)^2} 
 \nonumber\\ && \hspace{50pt}
 \times\varphi \bar{\varphi} \hat{\varphi} \varphi' \tr(g\sigma_{\beta}\gb \sigma_{\alpha} \gh \sigma_{\mu}g' \sigma_{\nu})
 D_{\alpha \beta}(k) D_{\mu \nu} (q)
 \, .
\end{eqnarray}
Note that this diagram has to be taken twice, and we include this factor in its definition. 
Applying the identities (\ref{id1}), (\ref{id2}) (with $n=0$) inside the Dirac trace, we can rewrite 

\be
 {\cal L}^{3A} (f) = \frac{e^4  }{32 \pi^6}  \int_{0}^{\infty}dz d\zb d\zh dz' ~ \bar{\g} \g'\int \frac{d^2 q~ d^2 k~ d^2 p}{k^4 q^4}\varphi \bar{\varphi} \hat{\varphi} \varphi' 
\tr(g\K \pbdash \K \gh \q \slash p' \q)
\ee
More explicitly, the trace can be written as

\bear
\tr(g\K \pbdash \K \gh \q \slash p' \q) = I_1 + I_2
\label{atrace}
\ear
where 
\beqa
I_1 & = & m^2 \tr (\e^{\sigma_3 (z-\zh)} \K \pbdash \K \q \slash p' \q ) \nonumber \\
I_2 & = & - \g \hat{\g} \tr ( \p \K \pbdash \K \p \q \slash p' \q) \nonumber\\
\label{atraces}
\eeqa
%The third term, given by 
%
%\be
%I_3 = -im\left[ \g \tr ( \e^{\sigma_3 \zh} \q \slash p' \q \p \K \pbdash \K ) + \hat{\g}\tr( \e^{\sigma_3 z} \K \pbdash \K \p \q \slash p' \q  )  \right]
%\ee
%vanishes on account of the trace identities of the sigma matrices.
\noindent
Further, we rewrite the $p$-dependent part of the argument of the exponential by defining  

\be
\varphi \bar{\varphi} \hat{\varphi} \varphi' \equiv \Omega \e^{-a p^2-2( \bar t k + t' q)\cdot p } \e^{-(\bar t k^2 + t' q^2)},
\ee
where we abbreviated $t\equiv \tanh z /ef$, $t' \equiv \tanh z' /ef$, etc., and defined

\beqa
a &=& t + \bar t + \hat t + t' ,\quad \Omega = \frac{\g \hat{\g} \bar{\g} \g' }{ (ef)^4} \e^{- 2\kappa (z+ \zh + \zb + z')}.
\nonumber\\
\eeqa
The amplitude is then given by

\be
{\cal L}^{3A} (f) = \frac{e^4  }{32 \pi^6} \int Dz ~  \bar{\g} \g' \Omega \int \frac{d^2 k ~d^2 q}{ k^4 q^4}  \e^{-(\bar t k^2 + t'q^2)} \int d^2 p ~\e^{ -ap^2 -2( \bar t k + t' q)\cdot p } ~~ (I_1 + I_2 )
\ee
where $\int Dz$ here and in the following is a shorthand for the fourfold $z$ - integral. 

\noindent
Working out the traces (\ref{atraces}) and performing the gaussian $p$ - integration, one finds

\be
{\cal L}^{3A} (f) =
 \frac{e^4 }{32 \pi^6 } \int Dz~ \bar{\g} \g' \Omega (J_1 + J_2)
\ee
where

\beqa
J_1 &=& \frac{2 \pi m^2 \cosh(z-\zh)}{a^2} \int \frac{d^2 k ~ d^2 q}{k^4 q^4} \e^{-d_A k^2 - e_A q^2 + 2 g_A k\cdot q} 
\nonumber\\ 
&& \times
\Bigg\{ [ 2(k \cdot q)^2 -k^2 q^2] 
\bigl[1 - d_Ak^2 - e_A q^2  + 2 g_A k \cdot q\bigr]
%\nonumber\\
%&& \hspace{215pt}
 + (t + \hat{t} )  k^2 q^2 (k \cdot q)   \Bigg\} 
\label{J1}\\
J_2 &=& -  \frac{ 2\pi \g \hat{\g}}{a^5} \int \frac{d^2 k ~ d^2 q}{k^4 q^4} \e^{-d_A k^2 - e_A q^2 + 2 g_A k\cdot q}  \Bigg\{ [ (t'^4  - a t'^3) q^4  + 
(\bar{t}^4 - a \bar{t}^3 ) k^4 ]  [ 2(k \cdot q)^2 -k^2 q^2] \nonumber\\
&&+ [4  \bar{t}^3 t' - a \bar{t}^3 -3 a \bar{t}^2 t' + a^2 \bar{t}^2 ] k^4 q^2 (k \cdot q) + [4  \bar{t} t'^3 - a t'^3 -3 a \bar{t} t'^2 + a^2 t'^2 ] k^2 q^4 (k \cdot q) 
\nonumber\\
&& + [6  \bar{t}^2 t'^2 - 3 a \bar{t}^2 t' - 3 a \bar{t} t'^2 + 2 a^2 \bar t t' ] k^4 q^4 \Bigg\}  
\label{J2}
\eeqa
Here $J_{1,2}$ still correspond to $I_1,I_2$, and we have introduced the further abbreviations

\be
d_A \equiv \frac{\bar{t}(a-\bar{t})}{a}, \quad e_A\equiv \frac{t'(a-t')}{a}, \quad g_A \equiv \frac{\bar t t'}{a} \, .
\label{defdefA}
\ee
Note that $d_A \geq 0,~~e_A \geq 0$, as required for convergence, and also that $d_Ae_A- g_A^2 \geq 0$.

We now come to the integrals over the photon momenta $k,q$, and
considering the factors of $\frac{1}{k^4 q^4}$ in (\ref{J1}), (\ref{J2}), it is not obvious that they are IR finite.
As was already mentioned, indeed in a general covariant gauge here one would encounter spurious IR divergences.
However, this happens not to be the case in traceless gauge; the $k,q$ integrals of $J_{1,2}$ are completely finite,
and moreover elementary. In appendix \ref{appintegrals} we list all the $k,q$ - integrals needed to complete our calculation of
diagram A, as well as of diagram B below.  

After these integrations, considerable simplifications occur, leading to the following simple results
for $J_{1,2}$:

\beqa
J_1 &=& \frac{2\pi^3 m^2 \cosh (z-\zh) }{a^2} \nonumber\\
J_2 &=& - \frac{4 \pi^3 }{a^3 \cosh z \cosh \zh} \nonumber\\
\label{J12fin}
\eeqa
Adding up both contributions and taking prefactors into account gives finally the integral representation:

\beqa
{\cal L}^{3A} (f) &=&  \frac{e^4}{16\pi^3 (ef)^4} \int_{0}^{\infty}dz d\zb d\zh dz' ~  \frac{\e^{- 2\kappa (z+ \zh + \zb + z')}}{a^2 \cosh z \cosh \zh (\cosh \zb \cosh z')^2 } \nonumber \\ 
& \times & \Bigg[ m^2 \cosh(z-\zh) - \frac{2  }{a \cosh z \cosh \zh} \Bigg] \nonumber\\
&=& \frac{e^4}{8 \pi^3 ef} \int_{0}^{\infty}dz d\zb d\zh dz' ~ \frac{\e^{- 2\kappa (z+ \zh + \zb + z')}}{A^2 \cosh z \cosh \zh (\cosh \zb \cosh z')^2} \nonumber \\
& \times & \Bigg[ \kappa \cosh (z-\zh)  - \frac{1}{A \cosh z  \cosh \zh} \Bigg]
\label{Afinal}
\eeqa
where we have introduced

\be 
A \equiv \tanh z + \tanh z' + \tanh \hat z + \tanh \bar z \, .
\label{defA}
\ee

\subsection{Diagram B}

\begin{figure}[h]
\begin{center}
\includegraphics[angle=0, width=0.3\textwidth]{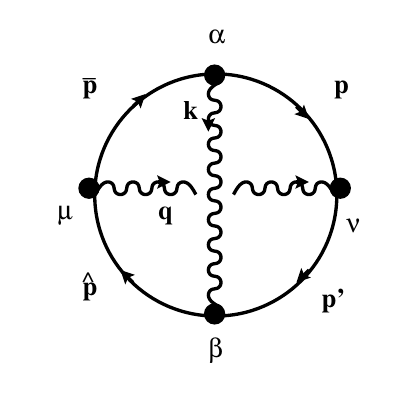}
\end{center}
\caption{Parametrization of diagram B.}
\label{figB}
\end{figure}

We come to diagram B, which is expected to be more difficult. See figure \ref{figB} for our parametrization.

\no
We again use $k$, $q$, and $p$ as the independent variables. The remaining variables are expressed in terms of them as

\bear
p' &=& p + q \non\\
\bar p &=& p+ k \non\\
\hat p &=& p + q + k \non\\
\label{kinrelB}
\ear
With these conventions, the contribution of this diagram is written as

\bear
{\cal L}^{3B} (f)
% &=& - \frac{e^4}{4} \int \frac{d^2 p}{(2 \pi)^2}\int \frac{d^2 k}{(2 \pi)^2}\int \frac{d^2 q}{(2 \pi)^2}
% \tr [G(p) \sigma_\alpha G(\bar p)\sigma_{\mu}G(\hat p)\sigma_{\beta}G(p')\sigma_{\nu}] D_{\alpha \beta}(k)D_{\mu\nu}(q)
%\nonumber\\
 &=&  - \frac{e^4}{4} \int Dz
\int\frac{d^2 p}{(2 \pi)^2}\int \frac{d^2 k}{(2 \pi)^2}\int \frac{d^2 q}{(2 \pi)^2}
\varphi\varphi' \bar\varphi \hat\varphi
 \, \tr [g \sigma_\alpha \bar g\sigma_{\mu} \hat g \sigma_{\beta}g'\sigma_{\nu}] D_{\alpha \beta}(k)D_{\mu\nu}(q)
 \nonumber\\
 \label{Bexplicit}
\ear
The Dirac trace in (\ref{Bexplicit}) can again be simplified using (\ref{id1}), (\ref{id2}).
The result can be written as

\bear
\tr [g \sigma_\alpha \bar g\sigma_{\mu} \hat g \sigma_{\beta}g'\sigma_{\nu}] 
D_{\alpha \beta}(k)D_{\mu\nu}(q)
=
b_0m^0 + b_2m^2 + b_4m^4 
\label{restrace}
\ear
where

\bear
b_0 &=& 4\gamma\gamma'\hat\gamma\bar\gamma \frac{1}{k^4q^4} \tr [\slash p\slash k \slash{\bar  p}\slash q \slash{\hat p}\slash k \slash p' \slash q]
\non\\
b_2 &=& 0 \non\\
b_4 &=& \frac{8}{k^4q^4}\Bigl\lbrack \cosh(z- z' + \hat z - \bar z) \bigl(2 (k\cdot q)^2 - k^2q^2 \bigr)
+ 2i \sinh(z- z' + \hat z - \bar z )k\cdot q \,k\wedge q \Bigr\rbrack\non\\
\label{b}
\ear
Here the term in $b_4$ involving $\sinh$ can be dropped since it is odd under the exchange $(k\leftrightarrow q,   z' \leftrightarrow \bar z)$.

\no
For diagram B it turns out to be convenient to first perform the gaussian $p$ integral, without even working out the trace appearing in $b_0$. 
This integral appears in the form

\bear
\int d^2p\, \lbrace 1,p_{\alpha},p_{\alpha}p_{\beta},p_{\alpha}p_{\beta}p_{\gamma},p_{\alpha}p_{\beta}p_{\gamma}p_{\delta}\rbrace
\,\e^{-\frac{1}{ef} (\tanh z\, p^2 + \tanh \bar z \, (p+k)^2 + \tanh \hat z\, (p+k+q)^2 + \tanh z'\, (p+q)^2)}
\nonumber\\
\label{pint}
\ear
where only $b_0$ involves terms with numerators. One can then rewrite the exponent in (\ref{pint}) as

\bear
-(t p^2 + \bar{t}  (p+k)^2 +  \hat{t}  (p+k+q)^2 + t' (p+q)^2)
=
-a p^2 -2 b\cdot p - c
\label{rewritexpon}
\ear
with $a, t, \bar{t} \ldots $ defined as for diagram A, and

\bear
b &=& (\bar t + \hat t) k + (\hat t + t') q \non\\
c &=& \bar t k^2 + t' q^2 + \hat t (k+q)^2 \non\\
\label{abc}
\ear
%Thus one has, for example in the quadratic case,
%
%\bear
%\int d^2 p \, p_{\alpha}p_{\beta} \,\e^{-a p^2 -2 b\cdot p - c} = \frac{\pi}{a} \Bigl( \frac{\delta_{\alpha\beta}}{2a} + \frac{b_{\alpha}b_{\beta}}{a^2}\Bigr)
%\,\e^{\frac{b^2}{a}-c}
%\label{gaussp2}
%\ear
Any Kronecker - $\delta$ coming out of the $p$ integration will belong to $b_0$, and it will make the trace in $b_0$
vanish on account of the first of the identities (\ref{idsigma}). Therefore the effect of the $p$ integration can also be described as a replacement

\bear
\int d^2 p \, N(p) \,\e^{-a p^2 -2 b\cdot p - c} \to \frac{\pi}{a}
N\Bigl( p_{\alpha} \to - \frac{b_{\alpha}}{a}\Bigr)
\,\e^{\frac{b^2}{a}-c}
\label{gaussreplace}
\ear
Thus for the $b_0$ term we find

\bear
\int d^2 p \,
 \tr [\slash p\slash k \slash{\bar  p}\slash q \slash{\hat p}\slash k \slash p' \slash q]
  \,\e^{-a p^2 -2 b\cdot p - c} 
  =
 -  \tr \Bigl[\frac{\slash b}{a} \slash k \bigl(-\frac{\slash b}{a}+ \slash k\bigr)\slash q\bigl (-\frac{\slash b}{a} +\slash k+\slash q\bigr) 
 \slash k \bigl(-\frac{\slash b}{a} + \slash q\bigr) \slash q\Bigr]
  \frac{\pi}{a}
  \,\e^{\frac{b^2}{a}-c}
  \, .
  \nonumber\\
  \label{b0p}
\ear
We further abbreviate 

\bear
\frac{b}{a} \equiv b_k k + b_q q
\label{abbb}
\ear
and work out the trace:

\bear
\tr \Bigl[\frac{\slash b}{a} \slash k \bigl(-\frac{\slash b}{a}+ \slash k\bigr)\slash q\bigl (-\frac{\slash b}{a} +\slash k+\slash q\bigr) 
 \slash k \bigl(-\frac{\slash b}{a} + \slash q\bigr) \slash q\Bigr]
 &=& 2 \biggl\lbrace 
 \Bigl[
 b_k(1-b_k) + b_q(1-b_q) - 6 b_k(1-b_k)b_q(1-b_q)
 \Bigr] k^4q^4
 \non\\
 &&
 + b_k(1-b_k)(1-2b_k)(1-2b_q)k^4q^2 k\cdot q 
 \non\\
 &&
 + b_q(1-b_q)(1-2b_q)(1-2b_k)k^2q^4 k\cdot q 
 \non\\
 &&
 - \bigl[b_k^2(1-b_k)^2k^4 + b_q^2(1-b_q)^2q^4\bigr] 
 \bigl[2(k\cdot q)^2 - k^2q^2\bigr] 
 \biggr\rbrace
 \, .
 \non\\
\label{traceB}
\ear
The remaining $k,q$ integrals are again finite and elementary. We rewrite the new exponent in (\ref{b0p}) as

\bear
\frac{b^2}{a} - c \equiv -d_B k^2 -e_B q^2 +2g_B k\cdot q
\label{newexp}
\ear
where 

\bear
d_B &=& \frac{(t+t')(\bar t + \hat t)}{a} \nonumber\\
e_B &=& \frac{(t+ \bar t)(t' + \hat t)}{a}\nonumber\\
g_B &=& \frac{t'\bar t - t \hat t}{a} \nonumber\\
\label{degB}
\ear
again fulfilling $d_B \geq 0, e_B\geq 0, d_Be_B- g_B^2 > 0$. 

The five integrals arising from (\ref{traceB}) are included in appendix \ref{appintegrals}.
After simple manipulations, one finds the following result for the total momentum integral of $b_0$:

\bear
 \int \frac{d^2 p}{(2 \pi)^2}\int \frac{d^2 k}{(2 \pi)^2}\int \frac{d^2 q}{(2 \pi)^2}
 \,\e^{-[t p^2 + \bar t (p+k)^2 + \hat t (p+k+q)^2 + t'\, (p+q)^2]}
 \, b_0
 &=&
- \gamma\gamma'\hat\gamma\bar\gamma \frac{1}{8\pi^3} \frac{1}{(t+t'+\hat t + \bar t)^3}
\non\\ && \hspace{-100pt}\times
\frac{(t^2+\hat t^2)(t'+\bar t) + (t'^2+\bar t^2)(t+\hat t)}{tt'\hat t + tt'\bar t+t\hat t\bar t+ t'\hat t \bar t}
\label{totintb0}
\ear
For $b_4$ we need only a single integral, given in (\ref{int1}), and leading to

\bear
 \int \frac{d^2 p}{(2 \pi)^2}\int \frac{d^2 k}{(2 \pi)^2}\int \frac{d^2 q}{(2 \pi)^2}
 \,\e^{-[t p^2 + \bar t (p+k)^2 + \hat t (p+k+q)^2 + t'\, (p+q)^2]}
 \, b_4
 &=&
\frac{1}{8\pi^3} \frac{\cosh(z- z' + \hat z - \bar z)}{t+t'+\hat t + \bar t}
\non\\&& \hspace{-250pt} \times
\biggl\lbrace
1+ \frac{(t+t'+\hat t + \bar t)(tt'\hat t + tt'\bar t+t\hat t\bar t+ t'\hat t \bar t)}{(t\hat t - t'\bar t)^2}
\ln \Bigl\lbrack \frac{(t+t'+\hat t + \bar t)(tt'\hat t + tt'\bar t+t\hat t\bar t+ t'\hat t \bar t)}{(t+t')(t+\bar t)(\hat t+ t')(\hat t + \bar t)} \Bigr\rbrack
\biggr\rbrace
 \nonumber\\
\label{totintb4}
\ear
Putting things together, our final result for diagram B becomes

\bear
{\cal L}^{3B} (f) &=& \frac{1}{32\pi^3}\frac{e^4}{ef}\int_0^{\infty}dzdz'd\hat zd\bar z \,\e^{-2\kappa (z+z'+\hat z+\bar z)}
\biggl\lbrace
\frac{1}{\cosh^2 z \cosh^2 z' \cosh^2\hat z\cosh^2 \bar z}
\frac{B}{A^3C} \non\\
&&\hspace{150pt} - 4\kappa^2 \frac{\cosh(z- z' + \hat z - \bar z)}{\cosh z \cosh z' \cosh \hat z\cosh \bar z}
\Bigl\lbrack
\frac{1}{A} - \frac{C}{G^2}\ln \Bigl(1+ \frac{G^2}{AC}\Bigr)
\Bigr\rbrack
\biggr\rbrace\non\\
\label{Bfinal}
\ear
where

\bear
%A &=& \tanh z + \tanh z' + \tanh \hat z + \tanh \bar z \non
B &=& (\tanh^2 z + \tanh^2 \hat z)(\tanh z' + \tanh \bar z) +  (\tanh^2 z' + \tanh^2 \bar z)(\tanh z + \tanh \hat z) \non\\
C &=& \tanh z\tanh z' \tanh \hat z + \tanh z \tanh z' \tanh \bar z + \tanh z \tanh \hat z \tanh \bar z + \tanh z' \tanh \hat z \tanh \bar z \non\\
%E &=& (\tanh z + \tanh z') (\tanh z + \tanh \bar z)(\tanh \hat z + \tanh z' ) (\tanh \hat z + \tanh \bar z) \non
G &=& \tanh z \tanh \hat z - \tanh z' \tanh \bar z \non\\
\label{defBCG}
\ear

\section{Calculation of the weak-field expansion coefficients}
\label{sec:compint}
\renewcommand{\theequation}{4.\arabic{equation}}
\setcounter{equation}{0}

As was explained in the introduction, our main motivation for this calculation of the EHL in 2D is in the properties of the weak-field expansion.
For the one-loop and two-loop contributions, the expansion coefficients have already been obtained in closed form in \cite{81}.
In this chapter, we will develop algorithms for the extraction of the three-loop expansion coefficients from the integral representations
obtained in the previous two chapters.

In \eqref{defcn} we defined the weak field expansion coefficients at $l$ loops by

\bear
{\cal L}^{(l)}(\kappa) &=& \frac{m^2}{2\pi} \sum_{n=0}^{\infty}(-1)^{l-1}c^{(l)}_n (i\kappa)^{-2n} \, .
\nonumber
\ear
Thus at three loops, and now introducing, instead of $\kappa$, the expansion variable
$\rho \equiv\frac{1}{2\kappa}= \frac{ef}{m^2}$, we can write

\bear
{\cal L}^{(3)}(f) &=& \frac{m^2}{2\pi} \sum_{n=0}^{\infty}(-4)^n c^{(3)}_n \rho^{2n} \, .
\label{defcn3loop}
\ear
It will be convenient to replace the coefficients $c^{(3)}_n$ by a new set of coefficients $\Gamma_n$
through

\bear
4^n c^{(3)}_n &\equiv & \frac{\tilde\alpha^2}{64}\Gamma_n
\label{c3toGamma}
\ear
and to denote the contributions of the two diagrams to $\Gamma_n$ by $\Gamma_n^{A,B}$,
of their sum by $\Gamma_n^{A+B}$.

\subsection{Diagram A}

The case of diagram A is straightforward. Starting from our final representation (\ref{Afinal}), we rescale 
$z = \frac{ef}{m^2}w= \rho w$ etc. Then we can write 

\begin{eqnarray}
{\cal L}^{3A} (f)
&=& \frac{\tilde\alpha^2m^2}{32\pi} \int_{0}^{\infty}dw dw'  d\hat w d\bar w~ I_A ~ \e^{-a} \, ,\nonumber\\
I_A &=& \frac{\rho^3}{A^2 \cosh \rho w \cosh \rho \hat w (\cosh \rho \bar w \cosh \rho w')^2} \Bigg[ \frac{\cosh \rho (w-\hat w)}{2\rho}  - \frac{1}{A \cosh \rho w  \cosh \rho \hat w} \Bigg]
\nonumber\\
\label{IA}
\end{eqnarray}
where 

\bear
a \equiv w+ w' + \hat w + \bar w \, .
\label{defa}
\ear
We expand the integrand $I_A$ in powers of $\rho$,

\bear
I_A = \sum_{n=0}^{\infty} \alpha_n \rho^{2n}  \, .
\label{expandIA}
\ear
Since

\bear
A &=& \tanh \rho w + \tanh \rho w' + \tanh \rho \hat w + \tanh \rho \bar w = \rho a + O (\rho^2)
\label{Aw}
\ear
the coefficients $\alpha_n$ of this expansion can be decomposed into terms of the form

\bear
\alpha_n = \sum_{k}\frac{{P}_{nk}(w, w', \hat w, \bar w)}{a^k}
\label{polformA}
\ear
with $k\geq 2$ and polynomials $P_{nk}$. After exponentiating the denominator using a Feynman parameter,

\bear
\frac{1}{a^k} = \frac{1}{\Gamma (k)} \int_0^{\infty} d\lambda ~ \lambda^{k-1}\,\e^{-\lambda a} \, ,
\label{feynpar}
\ear
the integrand factorizes in integrals over $w, w', \hat w, \bar w$ that are elementary.
The integrand of the final $\lambda$ - integral consists of terms of the form 

\bear
\frac{{P} (\lambda)}{(\lambda+1)^m}
\label{polformlambda}
\ear
with integer numbers $m\geq 2$ and polynomial numerators $P(\lambda)$. Thus the
$\lambda$ - integral is elementary, too. This also makes it clear that the coefficients $\Gamma_n^{A}$ are all rational numbers.

For example, at lowest order we find simply

\bear
\alpha_0 = - \frac{1}{a^3} + \frac{1}{2a^2}
\label{alpha0} 
\ear
leading to 

\bear
 \int_{0}^{\infty}dw dw'  d\hat w d\bar w\alpha_0 \,\e^{-a} &=& 
 \int_0^{\infty} d\lambda \half (\lambda -\lambda^2)\int_{0}^{\infty}dw dw'  d\hat w d\bar w\,\e^{-(\lambda +1)(w +\bar w +\hat w + w') }
 \nonumber\\
 &=& \half \int_0^{\infty} d\lambda  \frac{\lambda (1 -\lambda)}{(\lambda + 1)^4}  = - \frac{1}{12}
\label{IA0}
\ear
and to $\Gamma_0^A = -\frac{1}{3}$. Using this method, we have found it easy to obtain the first twelve expansion coefficients for diagram A.

\subsection{Easy part of diagram B}

We proceed to the much more difficult case of the non-planar diagram B. Rewriting again
$z = \frac{ef}{m^2}w= \rho w$ etc.  (\ref{Bfinal}) becomes

\begin{eqnarray}
{\cal L}^{3B} (f)
&=& \frac{\tilde\alpha^2m^2}{128\pi} \int_{0}^{\infty}dw dw' d\hat w d\bar w ~ I_B ~ \e^{-a} \, ,\nonumber\\
I_B &=& I_B^{\rm easy} + I_B^{\rm hard} \, , \nonumber\\
\label{Bw}
\ear
where

\begin{eqnarray}
I_B^{\rm easy} &=&
\frac{\rho^3}{\cosh^2 \rho w \cosh^2 \rho w' \cosh^2 \rho \hat w\cosh^2 \rho \bar w}
\frac{B}{A^3C} \non\\
I_B^{\rm hard} &=& 
- \rho \frac{\cosh(\rho \tilde{w}) }{\cosh \rho w \cosh \rho w' \cosh \rho \hat w\cosh \rho \bar w}
\Bigl\lbrack
\frac{1}{A} - \frac{C}{G^2}\ln \Bigl(1+ \frac{G^2}{AC}\Bigr)
\Bigr\rbrack
\nonumber\\
\label{IB}
\end{eqnarray}
with a corresponding rewriting of the functions $A,B,C,G$ defined in (\ref{defA}), (\ref{defBCG}),
$a$ defined as in (\ref{defa}) and $\tilde{w} \equiv w- w' + \hat w - \bar w$. 

We will treat $I_B^{\rm easy}$ and $I_B^{\rm hard}$ separately, calling the corresponding contributions to $\Gamma_n^B$
$\Gamma_n^{B, easy}$ and $\Gamma_n^{B, hard}$. 

The integration of $I_B^{\rm easy}$ can still be done using an extension of the factorization method used for diagram A above.
Expanding $I_B^{\rm easy} $ in powers of $\rho$ yields

\bear
I_B^{\rm easy} = \sum_{n=0}^{\infty} \beta_n^{\rm easy} \rho^{2n} 
\label{expandIBeasy}
\ear
where the coefficients $\beta_n^{\rm easy}$ can be decomposed into terms of the form

\bear
\beta_n^{\rm easy} = \sum_{k,l}\frac{{Q}_n^{kl}(w, w', \hat w, \bar w)}{a^kc^l}
\label{polformBeasy}
\ear
with 

\bear
c \equiv ww'\hat w + ww'\bar w + w\hat w \bar w + w'\hat w\bar w \, ,
\label{c}
\ear
$k\geq 2$ and polynomials $Q_n^{kl}$. The factor $\frac{1}{a^k}$ is exponentiated as in  
\eqref{feynpar}. In the factor $\frac{1}{c^l}$ we first rewrite

\bear
c = w w' \hat w \bar w \bigl(\frac{1}{w}+ \frac{1}{w'} +\frac{1}{\hat w} + \frac{1}{\bar w}\bigr)
\label{rewritec}
\ear
and then exponentiate only the second factor:

\bear
\frac{1}{c^l} = \frac{1}{\bigl(w w' \hat w \bar w \bigr)^l} \frac{1}{\Gamma (l)} 
\int_0^{\infty} d\mu ~ \mu^{l-1}\,\e^{-\mu  \bigl(\frac{1}{w}+ \frac{1}{w'} +\frac{1}{\hat w} + \frac{1}{\bar w}\bigr)} \, ,
\label{feynparc}
\ear
After this, the integrand factorizes in integrals over $w, w', \hat w, \bar w$, and these integrals are of the
form  

\bear
\int_0^{\infty} dw \, w^k \,\e^{-(\lambda + 1)w  - \frac{\mu}{w}} &=&
2 \Bigl(\frac{\mu}{\lambda+1}\Bigr)^\frac{k+1}{2}K_{-k-1}\bigl(2\sqrt{\mu(\lambda +1)}\bigr) \, .
\ear
One is left with a double integral over  $\lambda$ and $\mu$,  involving a product of four Bessel functions
and rational factors. Remarkably, MATHEMATICA has an algorithm for the analytic evaluation of this type of integral \cite{adamchik}. 
The result is of the form 

\bear
r_1 + r_2 \zeta_3
\label{r1r2}
\ear
with rational numbers $r_{1,2}$. 
At lowest order one has

\bear
\beta_0^{\rm easy} &=&  \frac{Q_0^{31}}{a^3c}\, , \nonumber\\ 
Q_0^{31} &=& (w^2+\hat w^2)(w'+\bar w)+(w+\hat w)(w'^2+\bar w^2) \, ,
\label{beta0easy} 
\ear
leading to 

\bear
\int_{0}^{\infty}dw dw'  d\hat w d\bar w\beta^{\rm easy}_0 \,\e^{-a} &=& 
64\int_0^{\infty} d\lambda \int_0^{\infty}d\mu\, \lambda^2 \Bigl(\frac{\mu}{\lambda+1}\Bigr)^{\frac{3}{2}}\nonumber\\
&&\times K_0^2\bigl(2\sqrt{\mu(\lambda +1)}\bigr)K_1\bigl(2\sqrt{\mu(\lambda +1)}\bigr)K_2\bigl(2\sqrt{\mu(\lambda +1)}\bigr)
\nonumber\\
&\stackrel{\sy{\mu \to (\lambda + 1)\mu}}{=}& 
64\int_0^{\infty} d\lambda \frac{\lambda^2}{(\lambda + 1)^4}\int_0^{\infty}d\mu\,\mu^{\frac{3}{2}}
K_0^2\bigl(2\sqrt{\mu}\bigr)K_1\bigl(2\sqrt{\mu}\bigr)K_2\bigl(2\sqrt{\mu}\bigr)
\nonumber\\
&=& - \frac{1}{12} + \frac{7}{8} \zeta_3 \, .
\label{beta0easyfin}
\ear
At the next order we find polynomials $Q_1^{31}$, $Q_1^{41}$, $Q_1^{32}$, where, for example,

\bear
Q_1^{31} &=& -\frac{1}{3} \Bigl[ 2(w+\hat w)(w'^4+\bar w^4) + (w^2+\hat w^2)(w'^3+\bar w^3)+ (w^3+\hat w^3)(w'^2+\bar w^2)+2 (w^4+\hat w^4)(w'+\bar w) \Bigr] \, .
\nonumber\\
\ear
The full calculation is already too long to be presented here, so let us jump to the final result:

\bear
\int_{0}^{\infty}dw dw'  d\hat w d\bar w\beta^{\rm easy}_1 \,\e^{-a} &=& \frac{161}{960} - \frac{77}{128} \zeta_3 \, .
\label{beta1easyfin}
\ear

%%%%%%%%%%%%%%%%%%%%%%
\subsection{Hard part of diagram B}
\label{subsec:hard}

Finding a  method of closed-form evaluation for $I_B^{\rm hard}$ is more difficult. 
One of the difficulties with its integration is that straightforward attempts will
create spurious divergences at $G=0$. 
For obtaining a first integral we will
make essential use of the high symmetry of this diagram. Namely, while diagram A above has
only a $\mathbb{Z}_2\times\mathbb{Z}_2 $ invariance generated by the interchanges
$w \leftrightarrow \hat w$ and $w' \leftrightarrow \bar w$, diagram B additionally to those has the 
symmetry under pair exchange $(w,\hat w) \leftrightarrow (w',\bar w)$. The group generated by these
three reflections is the eight-element dihedral group $D_4$. 

\noindent
The expansion of $I_B^{\rm hard}$

\bear
I_B^{\rm hard} = \sum_{n=0}^{\infty} \beta^{\rm hard}_n \rho^{2n} 
\label{expandIBhard}
\ear
yields coefficients that can be decomposed in the following way: let us introduce, besides $a$ and $c$, now also

\bear
g &\equiv & w \hat w - w' \bar w \nonumber\\
h &\equiv & ac+g^2 \nonumber\\
\label{gh}
\ear
Let us further introduce the basis functions

\bear
U_{k,l} &\equiv& \frac{1}{g^k (h-g^2)^l} \nonumber\\
V_k &\equiv& \frac{\ln (\frac{1}{1-g^2/h})}{g^k } \nonumber\\
%   \frac{\ln (\frac{1}{1-g^2/h})}{g^k c^l}  
\label{defUV}
\ear
Then we can, using $ac=h-g^2$, write

\bear
\beta^{\rm hard}_n = \sum_{k,l} u^{k,l}_n U_{k,l} +  \sum_k v^k_n V_k
\label{betadecomp}
\ear
where the $u^{k,l}_n$ and  $v^k_n$ are polynomial functions of the $w,w',\hat w,\bar w$. Here in general the index $k$can take both even and odd
values. However, for reasons that will become clear below we shall eliminate the case of odd $k$, multiplying by a factor of $g$ in the numerator and the denominator wherever necessary. 

Further, the symmetries of diagram B, together with the natural appearance of the variable $\tilde w$,
motivate us to introduce a differential operator $\tilde d$, 

\bear
\tilde d & \equiv & \frac{\partial}{\partial w}- \frac{\partial}{\partial w'}+ \frac{\partial}{\partial \hat w} -\frac{\partial}{\partial \bar w} 
\label{deftilded}
\ear
which is in some sense conjugate to $\tilde w$. Its action on $\tilde w, a, c, g , h$ is simple:

\bear
\td \tilde w &=& 4, \quad
\td a = 0, \quad
\tilde d g = a, \quad
\tilde d c = -2 g, \quad
\tilde d h = 0 \ .
\label{algcga}
\ear
Given that $h$ is constant under the action of $\tilde d$, as far as $\tilde d$ is concerned
$U_{k,l}$ and $V_k$ are functions of the single variable $g$. Moreover, they  are functions simple enough to be integrated in closed
form, for any $k,l$ and an arbitrary number of times. Given that also $\tilde d^2 g = \tilde d a = 0$, 
we can use integration-by-parts with respect to the variable $g$ to write any term of the form appearing in (\ref{betadecomp}) as a total derivative with respect to $\tilde d$,
simply by repeatedly integrating the factors $U_{k,l}$ and $V_k$ and differentiating the polynomials $u^{k,l}_n, v^k_n$ until the latter vanish.
Denoting by ${\rm deg}$ the degree as a polynomial in $w, w', \hat w, \bar w$, and by, e.g., $U_{k,l}^{(-i)}$ the
$i$ - fold indefinite integral of $U_{k,l}$ in $g$, we can then explicitly write $\beta^{\rm hard}_n$ as a total derivative with respect to $\tilde d$:
$\beta^{\rm hard}_n = \tilde d \theta_n$, where

\bear
\theta_n = \sum_{k,l}  \sum_{i=0}^{{\rm deg}(u^{k,l}_n)}(-1)^iU_{k,l}^{(-i-1)}\tilde d^i u^{k,l}_n  
+ \sum_k \sum_{i=0}^{{\rm deg}(v^k_n)}(-1)^iV_k^{(-i-1)}\tilde d^i v^k_n 
\, .
\label{defthetan}
\ear
Thus we have now reduced the integrand to boundary terms:

\bear
\int_{0}^{\infty}dw dw' d\hat w  d\bar w \,\e^{-a} \beta^{\rm hard}_n &=& \int_{0}^{\infty}dw d\bar w d\hat w dw'
 \bigl( \frac{\partial}{\partial w}- \frac{\partial}{\partial w'}+ \frac{\partial}{\partial \hat w} -\frac{\partial}{\partial \bar w} \bigr )
  \,\e^{-(w+w'+\hat w +\bar w)}
\theta_n
\, . \nonumber\\
\label{firstint}
\ear
Moreover, due to the exponential factor there are no contributions from the upper boundaries of the $w_i$ - integrals, and 
the four contributions from the lower boundaries must be all equal, due to the perfect symmetry of the graph $B$. Thus we can
choose to eliminate $\bar w$, and write

\bear
\int_{0}^{\infty}dw d w' d\hat w d\bar w \,\e^{-a} \beta^{\rm hard}_n &=&4\int_{0}^{\infty}dw dw' d\hat w \,\e^{-(w+w'+\hat w)}\, \theta_n\vert_{\bar w =0}
\, .
\label{reduced}
\ear
For the remaining three Feynman parameters, we introduce the global scaling variable $T = w+w'+\hat w$, and rescale
$w = T \alpha, w' = T \alpha', \hat w= T \hat \alpha$.  This leads to the form

\bear
\int_{0}^{\infty}dw d w' d\hat w d\bar w \,\e^{-a} \beta^{\rm hard}_n &=&4
\int_0^{\infty} dT T^2 \,\e^{-T} \int_0^1 d\alpha \int_0^1 d\alpha' \int_0^1 d\hat\alpha 
\,\delta(1-\alpha - \alpha' - \hat\alpha)
\, \theta_n\vert_{\bar w =0}
\, .
\nonumber
\label{reduced1}
\ear
After this rescaling, $\theta_n\vert_{\bar w =0}$ depends on $T$ only through a global factor
of $T^{2n}$:

\bear
 \theta_n\vert_{\bar w =0} = T^{2n} \omega_n (\alpha,\alpha',\hat\alpha)
 \, .
\label{thetatoomega}
\ear
Thus the $T$ - integral factors out, and we get

\bear
\int_{0}^{\infty}dw d w' d\hat w d\bar w \,\e^{-a} \beta^{\rm hard}_n &=&4 (2n+2)!
 \int_0^1 d\alpha \int_0^1 d\alpha' \int_0^1 d\hat\alpha 
\,\delta(1-\alpha - \alpha' - \hat\alpha)
\, \omega_n (\alpha,\alpha',\hat\alpha)
\, .
\nonumber
\label{doT}
\ear
We use the delta function to eliminate $\alpha'$, rather than $\alpha$, or $\hat\alpha$, since by the symmetry of
diagram $B$ the resulting integrand will be symmetric in $\alpha,\hat\alpha$. 

\bear
\int_{0}^{\infty}dw d w' d\hat w d\bar w \,\e^{-a} \beta^{\rm hard}_n &=&
4 (2n+2)! \int_0^1 d\alpha \int_0^{1-\alpha} d\hat\alpha f_n(\alpha,\hat\alpha)
\nonumber\\
\label{elalphaprime}
\ear
where we have renamed $f_n(\alpha,\hat\alpha) \equiv  \omega_n (\alpha,1-\alpha-\hat\alpha,\hat\alpha)$.
We note that at this stage we have the correspondences

\bear
a &=& 1 \nonumber\\
c &=& \alpha\hat\alpha(1-\alpha-\hat\alpha) \nonumber\\
g &=& \alpha\hat\alpha \nonumber\\
h &=& \alpha(1-\alpha)\hat\alpha(1-\hat\alpha) \nonumber\\
\label{acghfin}
\ear
For the computation of the final integral over $\alpha$ and $\hat\alpha$ in (\ref{elalphaprime}) it will
be advantageous to perform a change of variables from $\hat\alpha$ to $\Lambda$,

\bear
\Lambda \equiv \frac{g}{\sqrt{h}} = \sqrt{\frac{\alpha\hat\alpha}{(1-\alpha)(1-\hat\alpha)}}
\label{defLambda}
\ear
which yields

\bear
\int_0^1 d\alpha \int_0^{1-\alpha} d\hat\alpha f_n(\alpha,\hat\alpha)
&=& 2\int_0^1d\Lambda \Lambda \int_0^1d\alpha \frac{\alpha(1-\alpha)}{\bigl[\alpha+(1-\alpha)\Lambda^2]^2} f_n(\alpha,\hat\alpha(\alpha,\Lambda))
\, .
\nonumber\\
\label{finalintB}
\ear

Let us carry through this algorithm for the lowest coefficient $n=0$. Taking the $\rho \to 0$ limit in \eqref{IB} 
and eliminating $c$ through $c = (h-g^2)/a$ yields

\bear
\beta^{\rm hard}_0 &=&  -\frac{1}{a} + \frac{c}{g^2} \ln \left(1 + \frac{g^2}{ac} \right)
\nonumber\\
&=& -\frac{1}{a} + \Bigl(\frac{h}{ag^2} - \frac{1}{a}\Bigr)  
\ln \Bigl(\frac{1}{1-g^2/h}\Bigr)
\nonumber\\
&=& u_0^{0,0}U_{0,0} + v_0^0V_0 + v_0^2V_2
\label{beta0hard}
\ear
where

\bear
u_0^{0,0} &=& - \frac{1}{a} \nonumber\\
v_0^0 &=& -\frac{1}{a} \nonumber\\
v_0^2 &=& \frac{h}{a} \nonumber\\
\label{coeffbeta0}
\ear
For constructing $\theta_0$ via (\ref{defthetan}) we need

\bear
U_{0,0}^{(-1)} &=& g \nonumber\\
V_0^{(-1)} &=& 2g - 2\sqrt{h}\, {\rm arcth} \Bigl(\frac{g}{\sqrt{h}}\Bigr) + g \ln \Bigl(\frac{1}{1-g^2/h}\Bigr)\nonumber\\
V_2^{(-1)} &=& \frac{2}{\sqrt{h}}\, {\rm arcth} \Bigl(\frac{g}{\sqrt{h}}\Bigr) - \frac{1}{g} \ln \Bigl(\frac{1}{1-g^2/h}\Bigr)\nonumber\\
\label{indefint}
\ear
Going through the steps leading to (\ref{elalphaprime}), we get

\bear
f_0(\alpha,\hat\alpha) &=& \biggl\lbrack -3g+4\sqrt{h}\,{\rm arcth}\Bigl(\frac{g}{\sqrt{h}}\Bigr) - g(1+h/g^2) \ln \Bigl(\frac{1}{1-g^2/h}\Bigr) \biggr\rbrack
\nonumber\\
\label{f0}
\ear
with $g,h$ as in (\ref{acghfin}). After the change of variables (\ref{defLambda}) this becomes

\bear
f_0(\alpha,\Lambda)&=& \sqrt{h}\biggl\lbrack -3\Lambda 
+4\,{\rm arcth}\Lambda - (\Lambda+1/\Lambda) \ln \Bigl(\frac{1}{1-\Lambda^2}\Bigr) \biggr\rbrack
\label{f02}
\ear
where now

\bear
\sqrt{h} = \frac{\alpha(1-\alpha)\Lambda}{\alpha + (1-\alpha)\Lambda^2}
\label{sqrth}
\ear
Both integrals can be done in closed form, and one finds

\bear
2\int_0^1d\Lambda \Lambda \int_0^1d\alpha \frac{\alpha(1-\alpha)}{\bigl[\alpha+(1-\alpha)\Lambda^2]^2} f_0(\alpha,\Lambda)
= -\frac{13}{96} + \frac{7}{64}\zeta_3
\, .
\nonumber\\
\label{intalphaLambdafinal}
\ear
Thus in total we have shown

\bear
\int_{0}^{\infty}dw d w' d\hat w d\bar w \,\e^{-a} \beta^{\rm hard}_0 &=& 
= -\frac{13}{12} + \frac{7}{8}\zeta_3
\, .
\nonumber\\
\label{beta0final}
\ear
In principle, this algorithm after computerization can be used to calculate the weak-field expansion coefficients of $I_B^{\rm hard}$ to arbitrary order. 
The problem is that the polynomials $u^{k,l}_n$ and  $v^k_n$ rapidly grow in size with increasing $n$. However, since the complete integrand
is invariant under the action of the group $D_4$, and both $h$ and $g^2$ possess this invariance, all those polynomials must be invariant, too
(note that this would not be quite true had we permitted odd powers of $g$ in the denominator, since $g$ itself is only a semi-invariant: it is invariant under 
$w \leftrightarrow \hat w$ and $w' \leftrightarrow \bar w$, but changes sign under the pair exchange $(w,\hat w) \leftrightarrow (w',\bar w)$). 
This suggests to improve the algorithm by using the representation theory of the group $D_4$ to rewrite those polynomials more compactly in terms of some basic polynomial invariants of $D_4$.

As we explain in detail in appendix \ref{app-invariants}, the application of the general theory of polynomial representations of finite groups (see, e.g., \cite{benson-book}) to our
case of the group $D_4$, acting on polynomials of four variables, shows that all our numerator
polynomials can be rewritten in the form 

$$\frac {P(a,\tilde w,v,j)}{a} $$

\noindent
where $P(a,\tilde w, v,j)$ is a polynomial in the four invariants (more precisely semi-invariants, see appendix \ref{app-invariants}) $a,\tilde w, v,j$.
Of those $a,\tilde w$ we have already seen, and the remaining two are defined by

\bear
v &=& (w+\hat w)(w'+\bar w) + 2(w\hat w + w' \bar w) \ ,\nonumber\\
j &=&( w - \hat w - w' + \bar w)(w -  \hat w +  w' - \bar w) \ .\nonumber
\ear
This choice of a basis of invariants is well-adapted to our integration-by-parts procedure, since
but for $\tilde w$ all get annihilated by $\tilde d$. 

\noindent
We are now ready to tackle the $n=1$ case. For $\beta^{\rm hard}_1$ one finds

\beqa
\beta^{\rm hard}_1
= u_1^{00} + u_1^{20} \frac 1 {g^2} + \ln\Big(1+ \frac {g^2}{ac}\Big)
\Bigg\{v_1^2 \frac 1 {g^2} + v_1^4 \frac 1 {g^4} \Bigg\} \ .\nonumber
\eeqa
with coefficient functions that in our new basis read

%\bear
%u^{00}_1 &=& \frac{\hat{w} w' + w \bar{w} + \bar{w} \hat{w} + ww' - \bar{w} w' - w \hat{w} }{a} - \frac{1}{3a^2} (w^3 + \bar{w}^3 + \hat{w}^3 + w'^3) \ , \non \\
%u^{10}_1 &=& \frac{1}{3a} \Big[ w \bar{w} \hat{w} (2 \bar w - w - \hat{w}) + w \hat{w} w' (2 w' - w -\hat{w}) - w \bar w w' (2 w - \bar w -w') - \bar w \hat{w} w' (2 \hat{w} -\bar w -w')  \Big]\ , \non \\
%v^2_1 &=& c (w \hat{w} + \bar w w' - w \bar w - \bar w \hat{w} - w w' - \hat w w') \non \\ 
%&-& \frac{1}{3} \Big[ w \bar w \hat{w}(w^2 + \bar{w}^2 + \hat{w}^2) + w \bar w w' (w^2 + \bar{w}^2 + w'^2) + w \hat{w} w' (w^2 + \hat{w}^2  + w'^2) + \bar w \hat{w} w' (\bar{w}^2 + \hat{w}^2 + w'^2)  \Big] \ , \non \\
%v^4_1 &=& \frac{1}{6}cg [w^3 \hat{w} + w \hat{w}^3 - \bar{w}^3 w' - \bar w w'^3] \ .\nonumber\\
%\ear
%This $v^3_1$ is not the same as ours under change of variables, because the change of variables will split this into a piece belonging to v^2_1 and other to v^3_1.
%
%
%\bear
%u^{00}_1 &=& \frac{8a^4 - 48 h + 3j^2 -6a j \tilde{w}  -21 a^2 \tilde{w}^2 }{48 a^3} \nonumber \\ 
%v_1^3 &=& -\frac{\tilde{w}}{384} (16 h - j^2 + 2 a j \tilde{w} - a^2 \tilde{w}^2)(16 h - j^2 + 2 a j \tilde{w} - a^2 \tilde{w}^2 - 16 a \mu) 
%\ear
\beqa
u_1^{00}&=&\frac  a 6 -\frac {\tilde w^2} {2a} -\frac c {a^2} \ ,
\nonumber\\
u_1^{20}&=& -{\frac {{a}^{5}}{96}}-{\frac {{a}^{3}{{\it \tilde w}}^{2}}{192}}+\frac{{a}^
{3}v}    {16}-{\frac {{a}{{\it \tilde w}}^{4}}{192}}-\frac{{a}{v}^{2}}{12}+\frac{
{a}{{\it \tilde w}}^{2}v}{48}+{\frac {{a}{j}^{2}}{192}}+{\frac {{{\it \tilde w}}
^{2}{j}^{2}}{64a}}\nonumber\\
&&+c \left( -\frac{{a}^{2}}{16}-\frac{{{\it \tilde w}}^{2}}{6}+\frac{v}3 
 \right) \ , \nonumber\\
 v_1^2 &=&{\frac {5\,{a}^{5}}{768}}-{\frac {a{j}^{2}}{192}}-{\frac {{a}^{3}{{
\it \tilde w}}^{2}}{384}}-\frac{{a}^{3}v}{32}+{\frac {a{{\it \tilde w}}^{4}}{768}}+
\frac{a{v}^{2}}{48}+{\frac {a{{\it \tilde w}}^{2}v}{96}}+c \left( -{\frac {13\,{a}^{2
}}{24}}+{\frac {7\,{{\it \tilde w}}^{2}}{24}}+\frac{5v}6 \right) \ ,\nonumber\\
v_1^4&=&
{
\frac {7\,{a}^{7}{{\it \tilde w}}^{2}}{3072}}-{\frac {7\,{a}^{7}v}{768}}+{
\frac {5\,{a}^{5}{{\it \tilde w}}^{4}}{3072}}+{\frac {5\,{a}^{5}{v}^{2}}{192
}}+{\frac {{a}^{3}{{\it \tilde w}}^{6}}{3072}}-\frac{{a}^{3}{v}^{3}}{48}+{\frac {
av{j}^{2}{{\it \tilde w}}^{2}}{192}}-{\frac {{a}^{3}{{\it \tilde w}}^{4}v}{256}}-{
\frac {{a}^{3}{j}^{2}{{\it \tilde w}}^{2}}{256}}\nonumber\\
&&-{\frac {a{{\it \tilde w}}^{4}{j}^
{2}}{768}}+{\frac {{a}^{3}{{\it \tilde w}}^{2}{v}^{2}}{64}}-{\frac {5\,{a}^{
5}{{\it \tilde w}}^{2}v}{384}}+{\frac {{a}^{9}}{1024}}\nonumber\\
&&+c\left( {\frac {{a}^{6}}{64}}-\frac{{a}^{2}{{\it \tilde w}}^{2}v}{24}-\frac{{j}^{
2}v}{24}+\frac{{a}^{2}{v}^{2}}{12}+\frac{{a}^{2}{j}^{2}}{48}+\frac{{a}^{4}{{\it \tilde w}}^
{2}}{32}-
\frac{{a}^{4}v}{12}+{\frac {{a}^{2}{{\it \tilde w}}^{4}}{64}} \right) \ .\nonumber
\eeqa
It remains now to use \eqref{eq:csub} to eliminate $c$.

The intermediate expressions involved in the integration of $\beta^{\rm hard}_1$ are cumbersome, but after following through the algorithm we obtain the final result:

\bear
\int_{0}^{\infty}dw d w' d\hat w d\bar w \,\e^{-a} \beta^{\rm hard}_1 &=& 
 \frac{121}{64} - \frac{203}{128}  \zeta_3 \ .
\nonumber\\
\label{beta1final}
\ear
As a final remark, at high orders it might become useful to give a separate treatment to the 
denominator $\cosh (\rho w)\cosh (\rho w')\cosh (\rho \wh )\cosh (\rho \wb)$ since it has full $S_4$ symmetry, while the rest has only $D_4$ symmetry;  

We remark that this invariant-based approach could as well be applied to $I_B^{\rm easy}$, but it would require a higher-order calculation to see whether this would be more efficient than the method described above. 

\section{Low-order results}
\label{sec:loworder}
\renewcommand{\theequation}{5.\arabic{equation}}
\setcounter{equation}{0}

In Table \ref{t3} we give the first six coefficients for both diagrams A and B. The coefficients for A and the first two coefficients for B were
calculated analytically using the methods developed in the previous section. The remaining coefficients for B were obtained by numerical integration. 
Note that the coefficients for diagram A are rational, for B of the form $r_1 + r_2 \zeta_3$ with rational numbers $r_1,r_2$.

%
%OLD:
%
%\begin{table*}[h]
%\caption[t3]{The cutoff-independent part $\Gamma^{(3)}_0(n)$ of the 
%three-loop Euler-Heisenberg coefficients $c^{(3)}(n)$. The coefficients
%are given numerically, separately for diagram A (second column), diagram B (third column) and C (fifth column). The fourth column gives the 
%sum of the coefficients of A and B.
%}
%\vspace{20pt}
%\begin{tabular}{|c|cccc}
%\hline
%$n$ & $\Gamma^{(3A)}_0(n)$ & $\Gamma^{(3B)}_0(n)$ & $\Gamma^{(3(A+B))}_0(n)$ &$\Gamma^{(C)}(n)$ \\  \hline
%0  & $\frac{22}{27} $ & $-\frac{125}{54}+\frac{7}{4}\zeta_3$ &  $~~0.6036=-\frac{3}{2}+\frac{7}{4}\zeta_3$ & $0.150896$  \\\\
%1  & $-0.584$ &$-0.389 $  & $-0.973 $ & $-0.07058$ \\\\
%2 & $ 0.85$ & $2.13$ & $2.98$ & $0.11992$  \\\\
%3 & $-3.8$ & $-15.9$  & $-19.7$ & $-0.41646$\\\\ 
%4 & $40.0$ & $192$ &$232$ & $2.484$ \\\\
%5 & $-685 $  & $-3370 $  &$-4055$ &$-23.09$  \\\\
%6 & & & &$311.8$  \\\\
%7 & & & &$-5792$  \\\\
%8 & & & &$ 142122$  \\\\
%9 & & & &$-4 .425 \times 10^6$ \\\\
%10 & & & &$1.714 \times 10^8$  \\\\
%11 & & && $ -8.066 \times 10^9$  \\\\
%12 & & & &$ 4.53 \times 10^{11}$ \\\
%\end{tabular}
%\label{t3}
%\end{table*}
%

\begin{table*}[h]
\caption[t3]{The first six of the 
three-loop Euler-Heisenberg coefficients $c^{(3)}(n)$ given in terms of the numbers $\Gamma_n^{A,B}$. The coefficients
are given separately for diagram A (second column), ``easy part'' of diagram B (third column). ``hard part'' of B (fourth column),
total contribution of B (fifth column), and sum of A and B = quenched contribution (sixth column).
}
\vspace{20pt}
\begin{tabular}{|c|cccccc|}
\hline
$n$ & $\Gamma^A_n$ & $\Gamma^{B\rm easy}_n$ & $\Gamma^{B\rm hard}_n$ & $\Gamma^B_n$ & $\Gamma^{A+B}_n$ &
% &$\Gamma^C_n$ 
\\  \hline 
0  & $\phantom{\Bigl\vert}-\frac{1}{3}$ & $-\frac{1}{12}+\frac{7}{8}\zeta_3$ &  $-\frac{13}{12}+\frac{7}{8}\zeta_3$ & $-\frac{7}{6}+\frac{7}{4}\zeta_3$
 &  $-\frac{3}{2}+\frac{7}{4}\zeta_3$ &
% & $-\frac{3}{4} + \frac{7}{8}\zeta_3 = 0.301800$   
\\
&$ = - 0.333333$ & $= 0.968466$ & $= - 0.0315335$ & $= 0.936933$ &   $ = 0.6036$ &
\\ \hline
1  & $\phantom{\Bigl\vert} -\frac{1}{30}$ & $-\frac{161}{960} + \frac{77}{128}\zeta_3 $ &  $- \frac{121}{64} + \frac{203}{128}  \zeta_3 $ & $-\frac{247}{120} + \frac{35}{16}\zeta_3 $  & $-\frac{251}{120} + \frac{35}{16}\zeta_3 $ &
% &  $-\frac{169}{144} + \frac{35}{32} \zeta_3 = 0.141139$  
\\
&$ = - 0.033333$ & $= 0.555404$ & $= 0.0157621$ & $= 0.571166$ & $=0.537833$ &
\\ \hline 
2  & $\phantom{\Bigl\vert}\frac{17}{63}= 0.269841$ & $1.16888$ &  $0.100512$ & $1.26939$  & $1.53924$ &
% & $-\frac{112423}{23040} + \frac{4361}{1024} \zeta_3 =  0.239836$  
\\ \hline 
3  & $\phantom{\Bigl\vert}\frac{251}{99}= 2.53535$ & $4.99381$ &  $0.685160$ & $5.67898$  & $8.21433$ &
% & $0.794914$  
\\ \hline 
4  & $\phantom{\Bigl\vert}\frac{407944}{15015}=27.1691$ & $35.9139$ &  $6.44966$ & $42.3637$ & $69.5327$ &
% & $4.73605$ 
\\ \hline 
5  & $\phantom{\Bigl\vert}\frac{26559976}{69615}=381.527$ & $392.363$ &  $84.4585$ & $476.822$  & $858.348$ &
\\ \hline 
% & $46.18$  
%\\\\
%6  & $\frac{715576448}{101745}$ & $6078.65$ &  $$ & $$  & $$ 
%% & $623.6$  
%\\\\
%7  & $\frac{286575166816}{1716099}$ & $$ &  $$ & $$  & $$ 
%% & $1.158 \times 10^4$  
%\\\\
%8  & $\frac{98372037638656}{19684665}$ & $$ &  $$ & $$  & $$ 
%% & $2.842 \times 10^5$  
%\\\\
%9  & $\frac{504453818412032}{2731365}$ & $$ &  $$ & $$  & $$ 
%% & $8.850 \times 10^6$  
%\\\\
%10  & $\frac{197707721191591936}{23881935}$ & $$ &  $$ & $$  & $$ 
%% & $3.428 \times 10^8$  
%\\\\
%11  & $\frac{305010508844770152448}{688422735}$ & $$ &  $$ & $$  & $$ 
%% & $1.613 \times 10^{10}$  
%\\\\
%12 & $\frac{2703229671702144698712064}{96792236541}$ & $$ &  $$ & $$  & $$ 
%% & $9.06 \times 10^{11}$  
%\\\\
%13  & $\frac{2710591846291057146506510336}{1322827232727}$ & $$ &  $$ & $$  & $$ 
%% & $$  
\end{tabular}
\label{t3}
\end{table*}

To the best of our knowledge, there are no results available in the literature that could be used to perform some
check on our results. However, let us mention that we have an internal check for the first coefficient, since for
this one we had obtained an analytical value already in a previous calculation that used Feynman gauge \cite{77}.
In Feynman gauge, we had found

\bear
\Gamma_0^{A(Feynman\, gauge)} = \frac{22}{27} , \quad \Gamma_0^{B(Feynman\, gauge)} = -\frac{125}{54} + \frac{7}{4}\zeta_3 \, , \quad
\label{gammafeynmangauge}
\ear
to be compared with our present result in traceless gauge, 

\bear
\Gamma_0^{A(traceless\, gauge)} = -\frac{1}{3} , \quad \Gamma_0^{B(traceless\,gauge)} = -\frac{7}{6}+ \frac{7}{4}\zeta_3 \, . \nonumber\\
\label{gammatracelessgauge}
\ear
We see that the sum agrees, $\Gamma_0^{A+B} = - \frac{3}{2} + \frac{7}{4}\zeta_3$, and that the rational part becomes gauge-independent only in the
sum over both diagrams, while the $\zeta_3$ - part can come only from the non-planar diagram in any gauge. 

Computing a  number of coefficients sufficient to address the issues related to the exponentiation conjecture
will require a more substantial computational effort, which we leave for future work.

\section{Conclusions and Outlook}
\label{sec:conclusions}
\renewcommand{\theequation}{6.\arabic{equation}}
\setcounter{equation}{0}

We have presented here the calculation of the three-loop correction to the Euler-Heisenberg Lagrangian in 1+1 dimensional
massive spinor QED. The calculation
has been performed in parallel using standard Feynman diagrams and the worldline formalism, treating the constant external field 
non-perturbatively in both cases. In both formalisms, the use of the ``traceless'' gauge $\xi = -1$ for the internal photons
turned out beneficial in making the IR finiteness of the effective Lagrangian manifest term-by-term. 
In the Feynman diagram approach, this gauge choice moreover led to substantive simplifications. 
Both methods led to four-parameter integral representations, although of a quite different structure. 
The Feynman parameter calculation results in relatively compact integrands, particularly for the non-planar diagram A, while
the worldline formalism yields a more extensive integrand, but has the advantage that this integrand holds for both the planar and
the non-planar sector. We have further used the worldline formalism for a recalculation of the two-loop EHL that is simpler than
the Feynman diagram calculation of \cite{81}. 

To the best of our knowledge, this constitutes the first calculation of a three-loop effective Lagrangian in quantum electrodynamics,
and also the first three-loop calculation in 1+1 QED (in fact even at the two-loop level the only calculations in 1+1 QED that we have been
able to find in the literature are the already mentioned studies by \cite{dunkra,krasnansky} for the Scalar QED case,  and the recent 
\cite{samsonov} on super QED (more effort seems to have gone into exploring the expansion in the mass, see \cite{adam-scattering}, \cite{adam-masspert} and refs. therein).

Based on the representation obtained in the Feynman diagram approach, we have developed computerizable algorithms for the analytic calculation
of the weak-field expansion coefficients that, in principle, work to arbitrary orders. For diagram B, our algorithm make use of the invariant theory of the symmetry
group of the graph, the dihedral group $D_4$, in a way that may be generalizable to other multiloop graphs and thus of independent interest.  
As to explicit computation, here we have been satisfied with a low-order test, leaving to future work the task of obtaining a sufficient number of coefficients
to confirm or refute the exponentiation conjecture, which is our main motivation for pushing this computation to the three-loop level.

Moreover, the methods developed here should also become useful in an eventual calculation of the three-loop EHL in four dimensions, particularly
for the self-dual case which is to some extent a ``doubling up'' of the two-dimensional case. 

\acknowledgments
We thank Gerry McKeon for early collaboration, and 
D. Broadhurst, A. Das, G.V. Dunne, R. Jackiw, D. Kreimer, E. Panzer, E. Rabinovici, M. Reuter, and V.I. Ritus for various discussions and/or correspondence. 
Special thanks to P. Baumann for many explanations on the theory of polynomial group invariants, and to James P. Edwards 
for performing some independent checks on our Feynman diagram calculations. 

C. S. thanks CONACYT for support through project Ciencias Basicas 2014 No. 242461, and the Kavli
Institute for Theoretical Physics (KITP) of the University of California, Santa Barbara, for hospitality during the program {\sl Frontiers of intense laser physics}. M. R. thanks the IFM, UMSNH for hospitality. 
This research was supported in part by the National Science Foundation under Grant No. NSF PHY11-25915. I. H. is thankful to the Theoretisch-Physikalisches Institut of the FSUJ for support while part of this work was completed, funding is acknowledged from CONACyT through the SNI program and PROMEP grant dsa103.5/16/10224. 
\appendix

\section{Conventions and formulas for Euclidean 1+1 QED}
\label{appconv}
\renewcommand{\theequation}{A.\arabic{equation}}
\setcounter{equation}{0}
\bigskip

\noindent
{\it Dirac equation:}

\begin{eqnarray}
\Bigl(\sigma_{\mu}(\partial_{\mu}-ieA_{\mu}) + m\Bigr) \psi &=& 0
\label{diraceq}
\end{eqnarray}
($\mu =1,2$).

\noindent
{\it Free electron propagator:}

\begin{eqnarray}
\frac{1}{i\slash{p} + m} &=& {-i\slash{p} +  m\over p^2 +m^2}
\label{freeelprop}
\end{eqnarray}
($\slash{p}=\sigma_{\mu}p_{\mu}$).

\noindent
{\it Photon propagator:}

\begin{eqnarray}
D_{\mu\nu}(k) = {\delta_{\mu\nu}\over k^2} - (1-\xi) \frac{k_{\mu}k_{\nu}}{k^4}
\label{phoprop}
\end{eqnarray}
($\xi = 1$ Feynman gauge, $\xi = -1$ traceless gauge).

\bigskip

\noindent
{\it Vertex:} 

\vspace{-20pt}

\begin{eqnarray}
ie\sigma_{\mu}
\label{vertex}
\end{eqnarray}

\noindent
{\it Field strength tensor:}

\vspace{-20pt}

\begin{eqnarray}
F &=& \left(\begin{array}{cc}0 & f \\-f & 0 \end{array}\right)
%\label{defF}
\end{eqnarray}

\noindent
{\it Fock-Schwinger gauge:}

\begin{eqnarray}
A_{\mu}(x) &=& -\frac{1}{2} F_{\mn} x_{\nu}
\label{FSgauge}
\end{eqnarray}

\noindent
{\it Electron propagator in a constant field in Fock-Schwinger gauge:}

\begin{eqnarray}
\bigl[\sigma_{\mu}(\partial_{\mu} -ieA_{\mu})+m \bigr] G(x-x') &=& \delta (x-x')
\label{defgx}
\end{eqnarray}

\begin{eqnarray}
G(p) &=& \int_0^{\infty} dT \e^{-T\bigl(m^2+{\tanh z\over z}p^2\bigr)}
{1\over\cosh z} \Bigl(m\e^{\sigma_3 z} - {i\slash{p}\over\cosh z}\Bigr)
\label{gp}\\
G(x) 
%&=& \int {d^2p\over (2\pi)^2} \e^{ip\cdot x}g(p) \label{gx}\\
&=&
{1\over 4\pi} \int_0^{\infty} {dT\over T} \e^{-m^2T}{z\over\sinh z}
\e^{-{z\over \tanh z}{x^2\over 4T}}
\Bigl(m\e^{\sigma_3 z} +{1\over 2T} {z\over \sinh z} \slash{x}\Bigr)
\end{eqnarray} 
($z=efT$).

We use the following straightforward identities:

\beqa \nn
\sigma_{\mu}\sigma_{\nu_1}\cdots \sigma_{\nu_{2n+1}}\sigma_{\mu}
&=& 0\\ \nn
\s_\al \s_\bt &=& \dl_{\al \bt} {\bf 1} + i \ep_{\al \bt} \s_3 \\ \nn
\s_3 \s_\al   &=& -\s_\al \s_3 =  i\ep_{\al \bt} \s_\bt \\ \nn
%(\e^{\s_2 a}x ) \cdot (\e^{\s_2 b} y) &=& x \cdot y \cosh (a-b) + i x \wedge y \sinh(a-b)  \\ \nn
%(\e^{\s_2 a}x ) \wedge (\e^{\s_2 b} y) &=& x \wedge y \cosh (a-b) - i x \cdot y \sinh(a-b)  
\nonumber\\
\label{idsigma}
\eeqa
($\ep_{12}=1$).

%vfill\eject\noindent

\section{Momentum integrals appearing in the worldline calculation}

\renewcommand{\theequation}{B.\arabic{equation}}
\setcounter{equation}{0}
\label{appB}

\bear%%%%%%%%%%%%
\frac{1}{(2\pi)^2}
\int \frac{d^2k}{k^2}
\int \frac{d^2l}{l^2}
\,\e^{- \half (k,l) {\cal M} (k,l)}
k^2l^2
&=&
%\nonumber\\&& \hspace{-170pt}
\frac{1}{\Delta}
\nonumber\\
%%%%%%%%%%%%
\frac{1}{(2\pi)^2}
\int \frac{d^2k}{k^2}
\int \frac{d^2l}{l^2}
\,\e^{- \half (k,l) {\cal M} (k,l)}
k_{\alpha}k_{\beta}l^2
&=&
%\nonumber\\&& \hspace{-170pt}
\frac{\delta_{\alpha\beta}}{2\Delta}
\nonumber\\
%%%%%%%%%%%%
\frac{1}{(2\pi)^2}
\int \frac{d^2k}{k^2}
\int \frac{d^2l}{l^2}
\,\e^{- \half (k,l) {\cal M} (k,l)}
k^2l_{\alpha}l_{\beta}
&=&
%\nonumber\\&& \hspace{-170pt}
\frac{\delta_{\alpha\beta}}{2\Delta}
\nonumber\\
%%%%%%%%%%%%
\frac{1}{(2\pi)^2}
\int \frac{d^2k}{k^2}
\int \frac{d^2l}{l^2}
\,\e^{- \half (k,l) {\cal M} (k,l)}
k_{\alpha}k_{\beta}k_{\gamma}l_{\delta}
&=&
%\nonumber\\&& \hspace{-170pt}
\frac{h_{3,1}}{8}
\Bigl(\delta_{\alpha\beta}{\cal C}_{\gamma\delta}+\delta_{\alpha\gamma}{\cal C}_{\beta\delta}+\delta_{\beta\gamma}{\cal C}_{\alpha\delta}\Bigr)
\nonumber\\
%%%%%%%%%%%%%%%%\frac{1}{(2\pi)^2}
\frac{1}{(2\pi)^2}
\int \frac{d^2k}{k^2}
\int \frac{d^2l}{l^2}
\,\e^{- \half (k,l) {\cal M} (k,l)}
k_{\alpha}k_{\beta}l_{\gamma}l_{\delta}
&=&
%\nonumber\\&& \hspace{-230pt}
\frac{1}{8}
\Bigl(h_{2,2}^{kk,ll}\delta_{\alpha\beta}{\delta}_{\gamma\delta}
+h_{2,2}^{kl,kl}({\cal C}_{\alpha\gamma}{\cal C}_{\beta\delta}
+{\cal C}_{\beta\gamma}{\cal C}_{\alpha\delta})\Bigr)
\nonumber\\
%%%%%%%%%%%%%%%%%%%\frac{1}{(2\pi)^2}
\frac{1}{(2\pi)^2}
\int \frac{d^2k}{k^2}
\int \frac{d^2l}{l^2}
\,\e^{- \half (k,l) {\cal M} (k,l)}
k_{\alpha}l_{\beta}l_{\gamma}l_{\delta}
&=&
%\nonumber\\&& \hspace{-170pt}
\frac{h_{1,3}}{8}
\Bigl({\cal C}_{\alpha\beta}\delta_{\gamma\delta}+{\cal C}_{\alpha\gamma}\delta_{\beta\delta}+{\cal C}_{\alpha\delta}\delta_{\beta\gamma}\Bigr)
\nonumber\\
%%%%%%%%%%%%%%
\label{doints}
\ear
$h_{1,3},h_{3,1},h_{2,2}^{kk,ll},h_{2,2}^{kl,kl}$ have been given in (\ref{defh}).

\vfill\eject\noindent

\section{Momentum integrals appearing in the calculation of diagrams A and B}
\label{appintegrals}
\renewcommand{\theequation}{C.\arabic{equation}}
\setcounter{equation}{0}

Here we list all the integrals over the photon momenta $k,q$ appearing in the calculations of diagrams A and B in section
\ref{sec:ABfeynman}. The parameters $d,e,f$ should be replaced by the ones defined in (\ref{defdefA}) for diagram A, and 
defined in (\ref{newexp}) for diagram B.

\begin{subequations}\label{integrals}
\begin{align}
\int \frac{d^2 k}{k^4} \int \frac{d^2 q}{q^4}  \e^{ - dk^2 -e q^2 +2g k\cdot q} [ 2(k \cdot q)^2 -k^2 q^2] &=\! \pi^2 \bigg(\! 1 + \Big( \frac{de}{g^2}-1 \Big)
\ln \Big( 1- \frac{g^2}{de} \Big)\! \bigg) \label{int1}  \\
\int \frac{d^2 k}{k^4} \int \frac{d^2 q}{q^4} \e^{ - dk^2 -e q^2 +2g k\cdot q} k^2 [ 2(k \cdot q)^2 -k^2 q^2] &= -\frac{\pi^2}{d} \Bigg( 1 + \frac{de}{g^2} 
\ln \Big( 1- \frac{g^2}{de} \Big) \Bigg)  \\
\int \frac{d^2 k}{k^4} \int \frac{d^2 q}{q^4} \e^{ - dk^2 -e q^2 +2g k\cdot q} q^2 [ 2(k \cdot q)^2 -k^2 q^2] &= -\frac{\pi^2}{e} \Bigg( 1 + \frac{de}{g^2} 
\ln \Big( 1- \frac{g^2}{de} \Big) \Bigg)  \\
\int \frac{d^2 k}{k^4} \int \frac{d^2 q}{q^4} \e^{ - dk^2 -e q^2 +2g k\cdot q} (k \cdot q) [ 2(k \cdot q)^2 -k^2 q^2] &= -\frac{\pi^2}{g} \Bigg( 1 + \frac{de}{g^2} 
\ln \Big( 1- \frac{g^2}{de} \Big) \Bigg)  \\
\int \frac{d^2 k}{k^4} \int \frac{d^2 q}{q^4} \e^{ - dk^2 -e q^2 +2g k\cdot q} k^2 q^2 (k \cdot q) &= -\frac{\pi^2}{g} \ln \Big( 1- \frac{g^2}{de} \Big) \\
\int \frac{d^2k}{k^4} \int \frac{d^2q}{q^4} \, k^4 q^4 \,\e^{-dk^2 -e q^2 +2g k\cdot q} &= \frac{\pi^2}{ed-g^2} \\
\int \frac{d^2k}{k^4} \int \frac{d^2q}{q^4} \, k^4 q^2 k\cdot q \,\e^{-dk^2 -e q^2 +2g k\cdot q} &= \pi^2 \frac{g}{d^2e} \frac{1}{1- \frac{g^2}{de}} \\
\int \frac{d^2k}{k^4} \int \frac{d^2q}{q^4} \, k^2 q^4 k\cdot q \,\e^{-dk^2 -e q^2 +2g k\cdot q} &= \pi^2 \frac{g}{de^2} \frac{1}{1- \frac{g^2}{de}} \\
\int \frac{d^2k}{k^4} \int \frac{d^2q}{q^4} \, k^4 \bigl[ 2(k\cdot q)^2 - k^2q^2\bigr] \,\e^{-dk^2 -e q^2 +2g k\cdot q} &= \pi^2 \frac{g^2}{d^3e} \frac{1}{1-\frac{g^2}{de}} \\
\int \frac{d^2k}{k^4} \int \frac{d^2q}{q^4} \, q^4 \bigl[ 2(k\cdot q)^2 - k^2q^2\bigr] \,\e^{-dk^2 -e q^2 +2g k\cdot q} &= \pi^2 \frac{g^2}{de^3} \frac{1}{1-\frac{g^2}{de}} 
\end{align}
\end{subequations}

\section{Invariants of the dihedral group $D_4$}
\label{app-invariants}
\renewcommand{\theequation}{D.\arabic{equation}}
\setcounter{equation}{0}

Given a finite group $G$ and a  real $n-$dimensional representation (the representation can in fact be also
complex) $\rho(G) =\Gamma \subset GL(n,\mathbb R)$.
The representation $\rho$ may or may not be  irreducible. The representation $\rho$ induces naturally an action
on the set of polynonials with $n$ variables $\mathbb R[x_1,\cdots,x_n]$. Denoting $X=(x_1,\cdots,x_n)^t$ an $n-$dimensional vector
of ${\mathbb R}^\n$, 
the action  of $g \in \Gamma$ on $X$ is denoted $g\cdot X$.
A polynomial $I \in \mathbb R[x_1,\cdots,x_n]$ is said to be
invariant if for any $g \in \Gamma$ we have
\beqa
g\cdot I(X) \equiv I(g\cdot X) = I(X) \ .
\eeqa
The set of polynomial invariants is denoted
\beqa
 \mathbb R[x_1,\cdots,x_n]^\Gamma = \Big\{ I \in \mathbb R[x_1,\cdots,x_n] \ \  \text{s.t.} \ \ g\cdot I(X) = I(X) \ \ \forall g \in \Gamma \Big\} \ . 
\eeqa
The number of linearly independent polynomial invariants of degree $k$ is given by the $k-$th coefficient of
the Molien series, around zero, of the generating function \cite{benson-book}
\beqa
\Phi(t) = \frac 1{|G|} \sum \limits _{g \in \Gamma} \frac 1{\det(1 - t g)} \ . 
\eeqa
All polynomial invariants can be expressed in terms of what is usually called primitive  and secondary invariants.
The number of primitive invariants is equal to the dimension of the representation space, {\it i.e.}, is equal to $n$ \cite{sturm-book,benson-book}, so we denote them by $P_1,\cdots,P_n$.
They are algebraically independent, which means that they do not satisfy any polynomial identity of the type $f(P_1,\cdots,P_n)=0$. 
Denote now by ${\cal R} = \mathbb R[P_1,\cdots, P_n]$ the subalgebra of  polynomial invariants generated by the primitive invariants. Equivalently ${\cal R}$ can be seen as the set
of polynomials in $P_1,\cdots,P_n$. Denote $d_1,\cdots,d_n$ the degree of $P_1,\cdots, P_n$ respectively.
The number of secondary polynomial invariants is equal to $m=d_1 \cdots d_n/|G|$ \cite{decker-book}. Denote now $S_1,\cdots,S_m$ the set
of secondary polynomial invariants. The set of all invariants $(P_1,\cdots,P_n,S_1,\cdots,S_m)$ are not any more algebraically independent
and consequently do satisfy algebraic relations which are called syzygies.
It turns out that the subalgebra of invariants $\mathbb R[x_1,\cdots,x_n]^\Gamma$ is a free ${\cal R}-$module with basis $(S_1,\cdots,S_m)$.
In particular this means that any invariant $I \in  \mathbb R[x_1,\cdots,x_n]^\Gamma$ can be uniquely written as
\beqa
I=\sum\limits_{i=1}^m  f_i(P_1,\cdots,P_n) S_i \ , 
\eeqa
where $f_i(P_1,\cdots,P_n), i=1,\cdots, m$ belongs to ${\cal R}$, {\it i.e.}, are polynomials in $(P_1,\cdots,P_n)$.

There exist many automated way to compute primary and secondary invariants
and we have used the Computer Algebra System for Polynomial Computations  called
SINGULAR \cite{decker-book}. Before considering our specific case recall that the polynomial invariants for the $n-$dimensional
representation of the group of
permutation $\Sigma_n$ acting on $n$ elements consist of the well-known
symmetric polynomials. This in particular means that the set of symmetric polynomials of degree $d=1,\cdots,n$ is the
set of primitive invariants and the only secondary invariant is $\sigma_0=1$. In particular taking the notations
of the paper we have $n=4$ and $x_1=w, x_2= \hat w, x_3 = w', x_4=\bar w$
and the symmetric polynomials in this case are given by
\beqa
\sigma_1 &=& w + \hat w + w' + \bar w \ , \nonumber \\
\sigma_2&=& w \hat w + w w' + w \bar w +\hat w  w' + \hat w \bar w + w' \bar w\ , \label{sigmapoly} \\
\sigma_3&=& w  \hat w  w'+ w  \hat  w  \bar w +w  w'  \bar w + w' \hat w \bar w \, ,\nonumber \\
\sigma_4&=& w \hat w w' \bar w \ . \nonumber
\eeqa
The   dihedral group $D_4$ is eight dimensional and the three generators in the four dimensional (reducible) representation are
given by their action on $(w,\hat w, w' ,\bar w)$
\beqa
g_1 &:& w \leftrightarrow  \hat w \ , \nonumber\\
g_2 &:&  w' \leftrightarrow  \bar w \ , \\
g_3 &:& (w,\hat w) \leftrightarrow (w', \bar w) \ . \nonumber
\eeqa
It is obvious to see that not all the $\sigma$s are polynomials invariants for $D_4$.
For the group $D_4$ and for our representation, the Molien series takes the form

\beqa
\Phi(t) &=& \frac 18 \Bigg(\frac 2{1-t^4}+\frac 2{(1-t)^3(1+t)} +\frac 3{(1-t^2)^2}+\frac 1{(1-t)^4} \Bigg)\\ &=&
1+t+3 t^2+4 t^3+8 t^4+10 t^5+16 t^6 + O(t^7) \ .\nonumber 
\eeqa
For instance this means that there are $16$ linearly independent polynomial invariants  of degree six.  Since we are considering a representation
of dimension four of $D_4$ we must find four primitive invariants. Using SINGULAR, we obtain the primitive invariants

\beqa
a&=&\sigma_1=  w + \hat w + w' + \bar w \ , \nonumber \\
\lambda&=&(w + \hat w)(w' + \bar w) \ , \\
\mu &=&w \hat w + w' \bar w  , \nonumber \\
\sigma_4&=& w \hat w w' \bar w \ , \nonumber
\eeqa
of degree $d_1=1, d_2= d_3=2,d_4=4$ respectively. Thus the number of secondary invariants is two. Using  SINGULAR,
the secondary invariant
are given by
\beqa
\sigma_0=1 \ , \quad
c=\sigma_3= w w' \hat w + w w'\bar w + w \bar w \hat w + w' \hat w \bar w  \ .
\eeqa
Consequently any invariant can be uniquely written in the form
\beqa
I= f_1(a,\lambda,\mu,\sigma_4) + f_2(a,\lambda,\mu,\sigma_4) c \, .
\eeqa
There is a single syzygy, which is a degree six polynomial equation:

\beqa
\lambda(\mu^2-4\sigma_4)+a^2\sigma_4-a\mu c+c^2= 0 \ , 
\label{syzygy}
\eeqa
the set of primitive invariants are $(a,\lambda,\mu,\sigma_4)$ and the set of secondary invariants are $(1,c)$. 
We will denote such a set by $(1:a,2:\lambda,2:\mu,4:\sigma_4;3:c)$, where the first entry denotes the degree of the polynomial and the semi-colon separates the primitive from the secondary invariants.
This set is however not optimal for our algorithm developed in section \ref{subsec:hard}. based on integration by parts with the operator

\beqa
\widetilde {\text{d}}= \frac{\partial}{\partial w}-  \frac{\partial}{\partial w'}-  \frac{\partial}{\partial \bar w}+  \frac{\partial}{\partial \hat w}  \ ,
\eeqa
since most of the invariants are not annihilated by $\tilde d$:

\beqa
&\widetilde {\text{d}}a=0\ , \ \  \widetilde {\text{d}} \lambda= 2 \tilde w \ , \ \ \widetilde {\text{d}}\mu= -\tilde w \ ,  \nonumber\\
&\widetilde {\text{d}}\sigma_4=-w \hat w w'-w \hat w \bar w +w w' \bar w + \hat w w' \bar w \ , \ \
\widetilde {\text{d}} c = -2(w \hat w - w' \bar w) \ . \nonumber\\
\eeqa
Further, note that  $\widetilde {\text{d}}$ itself is not  invariant under the action of the whole  group  $D_4$,
for instance it changes by a sign under the permutation $(w,\hat w) \leftrightarrow (w',\bar w)$. Consequently
if its action  on an invariant polynomial $P$  is not vanishing, the resulting polynomial
is not an invariant polynomial. In fact, in this case $\tilde d P $ is a semi-invariant polynomial. A semi-invariant polynomial is an invariant polynomial up to a phase (in our case a minus sign) for some of the transformations of $D_4$.
Thus by means of $\widetilde {\text{d}}$ one can associate to any invariant polynomial not in the kernel of $\widetilde {\text{d}}$, a semi-invariant polynomial. In our case,
this suggests to introduce the semi-invariants:
\beqa
   \tilde w&=& w+ \hat w  - w' -\bar w \, ,
   \nonumber\\
  g&=&  w \hat w - \bar w w' \, ,\nonumber\\
 j &=&( w - \hat w - w' + \bar w)( w -  \hat w +  w' - \bar w) \, ,\nonumber\\
  j'&=& ( w-\hat w)(\bar  w- w') \, . \nonumber\\
  \eeqa
 It is easy to check that the Jacobian of the transformation $(w, w', \bar w,\hat w) \to (\tilde w, g,j,j')$ is non-vanishing,  which implies that the semi-invariants $(\tilde w, g,j,j')$ are  algebraically independent.
Now since

\beqa
\widetilde {\text{d}} \tilde w =4 \ ,\ \  \widetilde {\text{d}} j= 0 \ , 
 \eeqa
this suggests to consider the new set of primitive invariants of degree respective $1,2,2,4$:
 $(1:a,2:v,2: \tilde w^2, 4:j^2)$, where we have further introduced $v=2 \lambda + \mu$,
which now are all but $\tilde w$ in the kernel of $\widetilde {\text{d}}$.
Again the Jacobian of the transformation $(w,w',\hat w,\bar w) \to  (a,v,\tilde w^2, j^2)$ 
is found not to vanish, so that
%$128 \tilde w j j'$
this new set does not satisfy any polynomial relation and thus constitutes an alternative set of primitive invariants, well-adapted to $\widetilde {\text{d}}$. 

Thus any invariant can now be written as

\beqa
\label{eq:PI}
I(w,w',\bar w,\hat w) = f(a,v,\tilde w^2, j^2) +  g(a,v,\tilde w^2, j^2) c \  ,
\eeqa
with some polynomials $f$ and $g$. In this new basis the syzygy is given by

\beqa
{c}^{2}+ \left(\frac{{a}^{3}}8+\frac{a{{\it \tilde w}}^{2}} 8-\frac{av} 2 \right) c+{
\frac {{a}^{6}}{256}}+{\frac {{a}^{4}{{\it \tilde w}}^{2}}{128}}-\frac{{a}^{
4}v}{32}-\frac{{a}^{2}v{{\it \tilde w}}^{2}}{32}-{\frac {{j}^{2}{{\it \tilde w}}^{2}}{64}}+{
\frac {{a}^{2}{{\it \tilde w}}^{4}}{256}}+\frac{{a}^{2}{v}^{2}}{16}
=0 \ . \nonumber\\
\eeqa
Since $\widetilde {\text{d}} c =-2g$ we can further refine the procedure. Introduce $h=ac + g^2$ which is in the kernel of $\widetilde {\text{d}}$ and 
using the relationships

\beqa
\label{eq:csub}
c=\frac{4a^2v +j^2-a^4 -16 g^2}{16 a} \ ,\ \ g=\frac{a\tilde w-j}{4} \ ,
\eeqa
we can eliminate first $c$ and then $g$ so that the polynomial in \eqref{eq:PI} reduces to

\beqa
\label{eq:final-exp}
I(w,w',\bar w,\hat w) &=& f(a,v,\tilde w^2, j^2) +  g(a,v,\tilde w^2, j^2) 
% {\color{red} \cancel{\frac{ h -\frac1 {16}( a \tilde w - j)^2}{a}}
  \frac{4a^2 v+j^2-a^4-(a\tilde w-j)^2}{16 a}
\nonumber\\
&=& \frac {P(a,\tilde w, v,j)}{a}  \ .
\eeqa
Here $P$ is a polynomial in the variables $(a,\tilde w,v,j)$ which are algebraically independent since the Jacobian
  \begin{eqnarray}
\frac{\partial(a,v,j,\tilde w)}{\partial(w,w',\bar w,\hat w)} = -32(\bar w -w')(w-\hat w) \ne 0\ . \nonumber
    \end{eqnarray}
  Since the polynomial $f$ and $g$ are uniquely defined, and since there are no relation between the variables $(a,\tilde w, v, j)$ the polynomial $P$ is also uniquely
  defined.

Note that in this ``basis'' $I$ has  a non-polynomial expression  because we divide by $a$.  
However this does not matter, since $a$ will eventually be replaced by unity anyway in the further procedure of evaluating diagram $B$ . The benefit is that now all
polynomial invariants are expressed in terms of $(a,\tilde w, v, j)$ which are all,  except $\tilde w$, in the kernel of $\widetilde {\text{d}}$.
This is our optimal ``basis''. Note that $(a, v)$ are invariant polynomials whereas $(\tilde w, j)$ are semi-invariants polynomials. Note also that more precisely $P$ depends on $\tilde w^2, j^2$ and $\tilde w  j$ which are invariant under the whole action of the group $D_4$.

%\vfill\eject


\begin{thebibliography}{99}

\bibitem{eulhei}
W. Heisenberg and H. Euler, Z. Phys. {\bf 98} (1936) 714.

\bibitem{ditgie-book}
W. Dittrich and H. Gies, 
{\it Probing the Quantum Vacuum. Perturbative Effective Action Approach in Quantum Electrodynamics and its Application},
Springer Tracts Mod. Phys. {\bf 166} (2000) 1.

\bibitem{dunnerev}
G.V. Dunne, Cont. Adv. in QCD, {\bf 478} (2002), hep-th/0207046.

\bibitem{itzzub-book}
C. Itzykson and J. Zuber, {\it Quantum Field Theory}, McGraw-Hill 1985.

\bibitem{56}
L. C. Martin, C. Schubert and V. M. Villanueva Sandoval, 
%``On the low-energy limit of the QED N - photon amplitudes'', 
Nucl. Phys. B {\bf 668} (2003) 335, arXiv:hepth/0301022. 

\bibitem{118}
James P. Edwards, A. Huet and C. Schubert,
%``On the Low Energy Limit of the QED N Photon Amplitudes: part 2'',
Nucl. Phys. B {\bf 935} (2018) 198-209, arXiv:1807.10697 [hep-th].
%DOI: 10.1016/j.nuclphysb.2018.07.026. 

\bibitem{schwinger51}
J. Schwinger, Phys. Rev. {\bf 82} (1951) 664.

\bibitem{ritus-ginzburg}
V. I. Ritus, ``The Lagrangian Function of an Intense Electromagnetic
Field'', in {\it Proc. Lebedev Phys. Inst.} Vol. {\bf 168}, {\it Issues
in Intense-field Quantum Electrodynamics}, V. I. Ginzburg, ed., (Nova
Science Pub., NY 1987).

\bibitem{sauter}
F. Sauter, Z. Phys. {\bf 69} (1931) 742.

\bibitem{37} 
G.V. Dunne and C. Schubert, 
%``Two-loop Euler-Heisenberg QED pair-production rate'', 
Nucl. Phys. B 564 (2000) 59, [hep-th/9907190. 

\bibitem{ritusspin}
V. I. Ritus, 
%``Lagrangian of an intense electromagnetic field and quantum electrodynamics at short distances'', 
Zh. Eksp. Teor. Fiz {\bf 69} (1975) 1517 [Sov. Phys. JETP {\bf 42} (1975) 774].

\bibitem{ditreu-book}
W. Dittrich and M. Reuter, {\it Effective Lagrangians in Quantum Electrodynamics}, Springer 1985.

\bibitem{18}
M. Reuter, M. G. Schmidt and C. Schubert, 
%``Constant external fields in gauge theory and the spin 0, 1/2, 1 path integrals'',  
Ann. Phys. (N.Y.) {\bf 259} (1997) 313, hep-th/9610191. 

\bibitem{24} 
D. Fliegner, M. Reuter, M. G. Schmidt, C. Schubert, 
%``The two-loop Euler-Heisenberg Lagrangian in dimensional renormalization'', 
Teor. Mat. Fiz. {\bf 113} 289 (1997) 
[Theor. Math. Phys. {\bf 113} 1442 (1997)] hep-th/9704194.

\bibitem{korsch}
B. K{\"o}rs and M.G. Schmidt, 
%``The effective two loop Euler-Heisenberg action for scalar and spinor QED in a general constant background field'', 
Eur. Phys. J. {\bf C 6} (1999) 175, 
hep-th/9803144.

\bibitem{66}
G.V. Dunne, A. Huet, D. Rivera and C. Schubert,
% ``Closed-form weak-field expansion of two-loop Euler-Heisenberg Lagrangians'',
JHEP {\bf 0611} 013 (2006), hep-th/0609065.

\bibitem{lebrit}
 S.L. Lebedev, V.I. Ritus,  Zh. Eksp. Teor. Fiz. {\bf 86} (1984) 408. 
 %[Sov. Phys. JETP 59 (1984) 237].

\bibitem{ritusmass} 
V.I. Ritus,
% ``Method of eigenfunctions and mass operator in quantum electrodynamics of a constant field'', 
Zh. Eksp. Teor. Fiz. {\bf 75} (1978) 1560 [JETP {\bf 48} (1978) 788].

\bibitem{afalma}
I.K. Affleck, O. Alvarez and N.S. Manton, 
%``Pair Production At Strong Coupling In Weak External Fields'',% 
Nucl. Phys. {\bf B 197} (1982) 509.

\bibitem{feynman50}
R.P. Feynman, Phys. Rev. {\bf 80} (1950) 440.

\bibitem{feynman51}
R.P. Feynman, Phys. Rev. {\bf 84} (1951) 108.

\bibitem{50} 
G.V. Dunne and C. Schubert, 
%``Closed-form two-loop Euler-Heisenberg Lagrangian in a self-dual background'', 
Phys. Lett. {\bf B526} 55 (2002) hep-th/0111134.

\bibitem{51}
G. V. Dunne and C. Schubert, 
%Two-loop self-dual Euler-Heisenberg Lagrangians. I: Real part and helicity amplitudes'',
JHEP 0208, 053 (2002) [arXiv:hep-th/0205004]. 

\bibitem{52}
G. V. Dunne and C. Schubert,
%``Two-loop self-dual Euler-Heisenberg Lagrangians. II: Imaginary part and Borel analysis'',
JHEP 0206, 042 (2002) [arXiv:hep-th/0205005]. 

\bibitem{dufish1} 
M.J. Duff and C.J. Isham, Phys. Lett. B {\bf 86}, 157 (1979).

\bibitem{dufish2} 
M.J. Duff and C.J. Isham, Nucl. Phys. B {\bf 162} (1980) 271.

\bibitem{60}
G.V. Dunne and C. Schubert, J. Phys.: Conf. Ser. {\bf 37} (2006) 59 [hep-th/0409021].

\bibitem{dyson} 
F.J. Dyson,
%``Divergence of perturbation theory in quantum electrodyanmics'', 
Phys. Rev. {\bf 85} (1952) 631. 

\bibitem{thooftborel} G. 't Hooft, ``Can we make sense out of `quantum 
chromodynamics'?'', in {\it The Whys of Subnuclear Physics}, ed. A. Zichichi,
(Plenum, New York, 1978); Reprinted in: G. 't Hooft, {\it Under the spell of the
gauge principle}, (World Scientific, Singapore, 1994).

\bibitem{111}
I. Huet, M. Rausch de Traubenberg and C. Schubert,
%``Asymptotic Behaviour of the QED Perturbation Serie'',
Adv. High Energy Phys. 2017 (2017) 6214341, arXiv:1707.07655 [hep-th].

\bibitem{ritusscal}
V. I. Ritus, 
%``Connection between strong-field quantum electrodynamics with short-distance quantum electrodynamics'',
Zh. Eksp. Teor. Fiz {\bf 73} (1977) 807 [Sov. Phys. JETP {\bf 46} (1977) 423].

\bibitem{weisskopf}
V. Weisskopf, K. Dan. Vidensk. Selsk. Mat. Fy. Medd. {\bf 14} (1936) 1.

\bibitem{blviwi} 
S. Blau, M. Visser and A. Wipf, 
%``Analytic results for the effective action'', 
Int. J. Mod. Phys.  {\bf A6} (1991) 5409.

\bibitem{gitgav}
S.~P.~Gavrilov and D.~M.~Gitman, Phys.\ Rev.\ D {\bf 53}, 7162 (1996) [hep-th/9603152].

\bibitem{dunkra}
G.V. Dunne and M. Krasnansky,
%`` `Background field integration-by-parts' and the connection between one-loop and two-loop Heisenberg-Euler effective actions'',
JHEP {\bf 0604}:020 (2006) [hep-th/0602216].

\bibitem{krasnansky}
M. Krasnansky,  Int. J. Mod. Phys. {\bf A 23} (2008) 5201 [hep-th/0607230].

\bibitem{81}
I. Huet, D.G.C. McKeon and C. Schubert,
%``Euler-Heisenberg lagrangians and asymptotic analysis in 1+1 QED. Part I: two-loop'',
JHEP {\bf 1012} (2010) 036, arXiv:1010.5315 [hep-th].

\bibitem{giekar}
H. Gies and F. Karbstein,
%``An Addendum to the Heisenberg-Euler effective action beyond one loop'',
{\it JHEP} {\bf 1703} (2017) 108;  arXiv:1612.07251 [hep-th].  

\bibitem{karbstein}
F. Karbstein,
%``Tadpole diagrams in constant electromagnetic fields''
JHEP {\bf 1710} (2017) 075,
%DOI: 10.1007/JHEP10(2017)075
arXiv:1709.03819 [hep-th]. 

\bibitem{112}
James P. Edwards and C. Schubert,
%``One-particle reducible contribution to the one-loop scalar propagator in a constant field '',
Nucl. Phys. B {\bf 923} (2017) 339, 
%DOI;10.1016/j.nuclphysb.2017.08.002, 
arXiv:1704.00482 [hep-th].

\bibitem{113}
N. Ahmadiniaz, F. Bastianelli, O. Corradini, James P. Edwards and C. Schubert,
%``One-particle reducible contribution to the one-loop spinor propagator in a constant field '',
Nucl. Phys. B {\bf 924} (2017) 377,
%DOI;10.1016/j.nuclphysb.2017.09.012, 
arXiv:1704.05040 [hep-th].

\bibitem{frgish-book}
E.S. Fradkin, D.M. Gitman and S.M. Shvartsman, {\it Quantum Electrodynamics with Unstable Vacuum}, Springer 1991.

\bibitem{77}
I. Huet, D.G.C. McKeon and C. Schubert, 
%``Three-loop Euler-Heisenberg Lagrangian and asymptotic analysis in 1+1 QED'',
Proc. of {\sl Ninth Conference on  QUANTUM FIELD THEORY UNDER THE INFLUENCE OF EXTERNAL CONDITIONS (QFEXT 
'09)}, Eds. K. A. Milton and M. Bordag, p. 505, World Scientific 2010, arXiv:0911.022 [hep-th].

\bibitem{85}
I. Huet, M. Rausch de Traubenberg and C. Schubert,
%``The Euler-Heisenberg Lagrangian beyond one loop'',
Proc. of {\sl Eleventh Conference on 
QUANTUM FIELD THEORY UNDER THE INFLUENCE OF EXTERNAL CONDITIONS (QFEXT '11)}, 
Eds. M. Asorey, M. Bordag, E. Elizalde,  
Int. J. Mod. Physics Conf. Series {\bf 14} (2012), 383, arXiv:1112.1049 [hep-th]. 

\bibitem{108}
I. Huet, M. Rausch de Traubenberg and C. Schubert,
%`` Multiloop EulerÐHeisenberg Lagrangians, Schwinger Pair Creation, and the Photon S-Matrix'',
Proc. of {\sl International Wokshop on Strong Field Problems in Quantum Theory, 06-11 June 2016, Tomsk},
Russ. Phys. J. {\bf 59} (2017) 11, 1746. 

\bibitem{119}
I. Huet, M. Rausch de Traubenberg and C. Schubert,
%`` Multiloop Euler-Heisenberg Lagrangians, Schwinger pair creation, and the QED N - photon amplitudes'',
{\it Proceedings of Science} (LL2018) 035, arXiv:1807.05938 [hep-th]. 

\bibitem{berkos-prl}
Z. Bern and D. A. Kosower, 
%``Efficient calculation of one loop QCD amplitudes'', 
Phys. Rev. Lett. {\bf 66} (1991) 1669.

\bibitem{berkos-npb}
Z. Bern and D. A. Kosower, 
% ``The computation of loop amplitudes in gauge theory'',  
Nucl. Phys. {\bf B 379} (1992) 451.

\bibitem{strassler1}
M. J. Strassler, 
%``Field theory without Feynman diagrams: One-loop effective actions'', 
Nucl. Phys. {\bf B 385} (1992) 145, hep-ph/9205205.

\bibitem{strassler2}
M.J. Strassler, 
%``Field theory without Feynman diagrams: a demonstration using actions induced by heavy particles'',
SLAC-PUB-5978 (1992) (unpublished).

\bibitem{shaisultanov}
R. Zh. Shaisultanov, 
%``On the string-inspired approach to QED in external field'', 
Phys. Lett. {\bf B 378} (1996) 354, hep-th/9512142.

\bibitem{40} 
C. Schubert,
% ``Vacuum polarisation tensors in constant electromagnetic fields: Part I'', 
Nucl. Phys. {\bf B 585}, 407 (2000), hep-ph/0001288.

\bibitem{41}
C. Schubert, 
%``Perturbative quantum field theory in the string-inspired formalism'', 
Phys. Rept. {\bf 355}, 73 (2001), arXiv:hep-th/0101036.  

\bibitem{91}
N. Ahmadiniaz, C. Schubert and V.M. Villanueva, 
%``String-inspired representations of photon/gluon amplitudes'',
%Preprint MaPhy-AvH/2012-08,
JHEP {\bf 1301}, 312 (2013), arXiv:1211.1821 [hep-th].

\bibitem{26}
C. Schubert,
%``The structure of the Bern-Kosower integrand for the N-gluon amplitude'',
Eur. Phys. J. C{\bf 5}, 693 (1998), hep-th/9710067. 

\bibitem{8} 
M. G. Schmidt and C. Schubert, 
%``Worldline Green Functions for Multiloop Diagrams'', 
Phys. Lett. B {\bf 331} 69 (1994),  hep-th/9403158.

\bibitem{15} 
M.G. Schmidt and C. Schubert, 
%``Multiloop calculations in the string-inspired formalism: the single spinor-loop in QED'', 
Phys. Rev. D 53, 2150 (1996), hep-th/9410100.

\bibitem{adamchik}
V. Adamchik,
%``The evaluation of integrals of Bessel functions via G-function identities'',
%https://doi.org/10.1016/0377-0427(95)00153-0
J. Comp. Appl. Math. {\bf 64} 3 (1995) 283. 

\bibitem{benson-book} L. C. Grove and C. T. Benson, ``Finite Reflection Groups", Springer Verlag (1985), Chapter 7.

\bibitem{samsonov}
I. B. Samsonov, 
%``Low-energy effective action in two-dimensional SQED: A two-loop analysis'',
JHEP 1707 (2017) 146, arXiv:1704.04148 [hep-th].

\bibitem{adam-scattering}
C. Adam,
%``Scattering processes in the massive Schwinger model'',
Phys. Rev. D {\bf 55} (1997) 6299,
%DOI: 10.1103/PhysRevD.55.6299
hep-th/9701013. 

\bibitem{adam-masspert}
C. Adam,
%``Massive Schwinger model within mass perturbation theory'',
Ann. Phys. {\bf 259} (1997) 1, hep-th/9704064. 
%DOI: 10.1006/aphy.1997.5697


\bibitem{sturm-book}
B. Sturmfels, {\it Algorithms in invariant theory}, 2nd ed., Springer 2008.
%DOI = {10.1007/978-3-211-77417-5},

\bibitem{decker-book}
W. Decker and C. Lossen, 
{\it Computing in algebraic geometry. A quick start using SINGULAR},
Springer 2006.

%\bibitem{oldham-book} 
%K. B. Oldham and J. Spanier, {\it The Fractional Calculus}, Academic Press Inc 1978, Chapter 2.




\end{thebibliography}
\end{document}